\documentclass{lmcs}
\pdfoutput=1

\usepackage{lastpage}
\renewcommand{\dOi}{14(4:23)2018}
\lmcsheading{}{1--\pageref{LastPage}}{}{}%
{May~29,~2017}{Dec.~11,~2018}{}

\usepackage{amsmath}
\usepackage{amssymb}
\usepackage{amsthm}
\usepackage{hyperref}
\usepackage{microtype}
\usepackage{marginnote}
\usepackage{stmaryrd}
\usepackage{tikz}
\usepackage[all,pdf]{xy}
\CompileMatrices 

\newcommand{\pushoutcorner}[1][dr]{\save*!/#1+1.4pc/#1:(1,-1)@^{|-}\restore}

\title{Axioms for Modelling Cubical Type Theory in a Topos}

\author[I. Orton]{Ian Orton}
\author[A.~M. Pitts]{Andrew M. Pitts}
\address{University of Cambridge, Department of Computer Science and Technology, UK}
\email{\{Ian.Orton,Andrew.Pitts\}@cl.cam.ac.uk}



\newcommand{\Bool}{\text{\normalfont\texttt{\{0{,}1\}}}}
\newcommand{\C}{\mathbf{C}}

\newcommand{\code}[2]{\ulcorner#1,#2\urcorner}
\newcommand{\coerce}{\kw{coerce}}
\newcommand{\Cof}{\kw{Cof}}
\DeclareMathOperator{\cof}{\kw{cof}}
\DeclareMathOperator{\colim}{colim}

\newcommand{\comp}{\circ}
\newcommand{\compat}{\smallsmile}
\DeclareMathOperator{\Comp}{\kw{Comp}}
\newcommand{\conc}{\mathbin{@}}
\renewcommand{\conj}{\wedge}
\DeclareMathOperator{\Contr}{\kw{Contr}}
\DeclareMathOperator{\Cpf}{\Box}
\newcommand{\D}{\mathbb{D}}
\renewcommand{\d}[1][x]{\mathsf{#1}}
\newcommand{\Dec}{\Omega_{\mathrm{dec}}}
\newcommand{\defeq}{\triangleq}

\newcommand{\disj}{\vee}
\newcommand{\dm}{\mathcal{C}}
\newcommand{\DM}{\mathbf{DM}}
\newcommand{\E}{\mathcal{E}}
\newcommand{\El}{\mathcal{E}\!\ell}
\DeclareMathOperator{\elim}{\kw{elim}}
\newcommand{\ent}{\vdash}
\DeclareMathOperator{\Equiv}{\kw{Equiv}}
\DeclareMathOperator{\equivToPath}{\kw{equivToPath}}
\DeclareMathOperator{\Ext}{\kw{Ext}}
\newcommand{\exto}{\nearrow}
\newcommand{\Face}{\mathbb{F}}
\newcommand{\False}{\bot}
\newcommand{\fdm}{\mathsf{dM}}

\DeclareMathOperator{\Fib}{\kw{Fib}}
\DeclareMathOperator{\fil}{\kw{fill}}
\DeclareMathOperator{\Fill}{\kw{Fill}}
\DeclareMathOperator{\fst}{\kw{fst}}

\newcommand{\fun}{\mathbin{\shortrightarrow}}
\DeclareMathOperator{\glue}{\kw{glue}}
\DeclareMathOperator{\Glue}{\kw{Glue}}
\newcommand{\I}{\kw{I}}
\newcommand{\iobj}{\kw{i}}
\newcommand{\id}{\mathit{id}}
\DeclareMathOperator{\idp}{\kw{k}}
\DeclareMathOperator{\Id}{\kw{Id}}
\newcommand{\IdU}[2]{#1 \path_{\Univ} #2}
\newcommand{\imp}{\Rightarrow}

\DeclareMathOperator{\inl}{\kw{inl}}
\DeclareMathOperator{\inr}{\kw{inr}}
\newcommand{\inv}[1]{\overline{#1}}
\DeclareMathOperator{\isFib}{\kw{isFib}}
\newcommand{\iso}{\cong}
\newcommand{\join}{\cup}
\newcommand{\kw}[1]{\mathtt{#1}}

\newcommand{\mono}{\rightarrowtail}
\newcommand{\morphism}{\rightarrow}
\newcommand{\Nat}{\kw{N}}
\newcommand{\Not}[1]{\tgt\mathbin{\textsf{-}}#1}
\DeclareMathOperator{\obj}{obj}
\newcommand{\op}{\mathrm{op}}
\newcommand{\Pair}[2]{\langle#1\mathrel{,}#2\rangle}
\renewcommand{\path}{\sim}
\DeclareMathOperator{\Path}{\kw{Path}}
\DeclareMathOperator{\pathToEquiv}{\kw{pathToEquiv}}

\newcommand{\pos}[1]{\kw{ax_{#1}}}

\newcommand{\Prop}{\Omega}

\DeclareMathOperator{\refl}{\kw{refl}}
\newcommand{\resby}{|}
\renewcommand{\S}{\mathcal{S}}

\newcommand{\Set}{\mathbf{Set}}
\DeclareMathOperator{\SGlue}{\kw{SGlue}}
\newcommand{\Simplex}{\mbox{\boldmath$\Delta$}}
\DeclareMathOperator{\snd}{\kw{snd}}
\newcommand{\src}{\kw{0}}

\newcommand{\sSet}{\hat{\Simplex}}
\newcommand{\strictify}{\pos{9}}
\DeclareMathOperator{\Succ}{\kw{S}}

\newcommand{\tgt}{\kw{1}}
\newcommand{\True}{\top}

\newcommand{\Univ}{\mathcal{U}}
\newcommand{\UnivFib}{\mathcal{V}}

\newcommand{\y}{\mathrm{y}}
\newcommand{\Zero}{\kw{Z}}

\begin{document}

\begin{abstract}
  The homotopical approach to intensional type theory views proofs of
  equality as paths. We explore what is required of an object $\I$ in
  a topos to give such a path-based model of type theory in which
  paths are just functions with domain $\I$.  Cohen, Coquand, Huber
  and M\"ortberg give such a model using a particular category of
  presheaves.  We investigate the extent to which their model
  construction can be expressed in the internal type theory of any
  topos and identify a collection of quite weak axioms for this
  purpose. This clarifies the definition and properties of the notion
  of uniform Kan filling that lies at the heart of their constructive
  interpretation of Voevodsky's univalence axiom.
\end{abstract}

\maketitle

\section{Introduction}
\label{sec:int}

Cubical type theory~\cite{CoquandT:cubttc} provides a constructive
justification of Voevodsky's univalence axiom, an axiom that has
important consequences for the formalisation of mathematics within
Martin-L\"of type theory~\cite{HoTT}. Working informally in
constructive set theory, Cohen~\emph{et al}~\cite{CoquandT:cubttc}
give a model of their type theory using the category $\hat{\dm}$ of
set-valued contravariant functors on a small category $\dm$ that is
the Lawvere theory for de~Morgan
algebra~\cite[Chapter~XI]{BalbesR:disl};
see~\cite{SpittersB:cubsct}. The representable functor on the generic
de~Morgan algebra in $\dm$ is used as an interval object $\I$ in
$\hat{\dm}$, with proofs of equality modelled by the corresponding
notion of path, that is, by morphisms with domain $\I$. Cohen~\emph{et
  al} call the objects of $\hat{\dm}$ \emph{cubical sets}. They have a
richer structure compared with previous, synonymous
notions~\cite{CoquandT:modttc,HuberS:modttc}. For one thing they allow
path types to be modelled simply by exponentials $X^\I$, rather than
by name abstractions~\cite[Chapter~4]{PittsAM:nomsns}. More
importantly, the de~Morgan algebra operations endow the interval $\I$
with structure that considerably simplifies the definition and
properties of the constructive notion of Kan filling that lies at the
heart of~\cite{CoquandT:cubttc}. In particular, the filling operation
is obtained from a simple special case that \emph{composes} a filling
at one end of the interval to a filling at the other end.
Coquand~\cite{CoquandT:cork} has suggested that this distinctive
composition operation can be understood in terms of the properties of
partial elements and their extension to total elements, within the
internal higher-order logic of toposes~\cite{LambekJ:inthoc}.  In this
paper we show that that is indeed the case and usefully so. In
particular, the \emph{uniformity} condition on composition
operations~\cite[Definition~13]{CoquandT:cubttc}, which allows one to
avoid the non-constructive aspects of the classical notion of Kan
filling~\cite{CoquandT:krimss}, becomes automatic when the operations
are formulated internally. Our approach has the usual benefit of
axiomatics -- helping to clarify exactly which properties of a topos
are sufficient to carry out each of the various constructions used to
model cubical type theory~\cite{CoquandT:cubttc}, clearing the way for
simplifications, generalisations and new examples.

To accomplish this, we find it helpful to work not in the higher-order
predicate logic of toposes, but in an extensional type theory equipped
with an impredicative universe of propositions $\Prop$, standing for
the subobject classifier of the topos~\cite{MaiettiME:modcdt}.
Working in such a language, our axiomatisation concerns two structures
that a topos $\E$ may possess: an object $\I$ that is endowed with
some elementary characteristics of the unit interval; and a subobject
of propositions $\Cof\mono\Prop$ whose elements we call
\emph{cofibrant propositions} and which determine the subobjects that
are relevant for a Kan-like notion of filling. (In the case
of~\cite{CoquandT:cubttc}, $\Cof$ classifies subobjects generated by
unions of faces of hypercubes.) Working internally with cofibrant
propositions rather than externally with a class of cofibrant
monomorphisms leads to an appealingly simple notion of
\emph{fibration} (Section~\ref{sec:cchmf}), with that of
Cohen~\emph{et al} as an instance when the topos is $\hat{\dm}$. These
fibrations are type-families equipped with extra structure
(\emph{composition} operations) which are supposed to model
intensional Martin-L\"of type theory, when organised as a Category
with Families (CwF)~\cite{DybjerP:intt}, say. In order that they do
so, we make a series of postulates about the interval and cofibrant
subobjects that are true of the presheaf model in
\cite{CoquandT:cubttc}. An overview of these postulates is given in
Section \ref{sec:axioms}, followed in subsequent sections by
constructions in the CwF of fibrations that show it to be a model of
Martin-L\"of type theory with $\Sigma$, $\Pi$, data types (we just
consider the natural numbers and disjoint unions) and intensional
identity types. In section~\ref{sec:glueing} we give an internal
treatment of \emph{glueing}, needed for constructions relating to
univalence~\cite{HoTT}. Our approach to glueing is a bit different
from that in~\cite{CoquandT:cubttc} and enables us to isolate a
strictness property of cofibrant propositions (axiom $\pos{9}$ in
Figure~\ref{fig:axi}) independently of glueing, as well as separating
out the use of the fact that cofibrant predicates are closed under
universal quantification over the interval (axiom $\pos{8}$ in
Figure~\ref{fig:axi}). In section~\ref{univalence} we give a result
about univalence that is provable from our axioms. However, for
reasons discussed at the end of that section, one cannot give an
internal account of the univalent universe construction
from~\cite{CoquandT:cubttc}. (This problem is circumvented
in~\cite{LicataD:intumhtt} by extending the internal type theory with
a suitable modality.)

In Section~\ref{sec:sata} we indicate why the model in
\cite{CoquandT:cubttc} satisfies our axioms and more generally which
other presheaf toposes satisfy them. There is some freedom in choosing
the subobject of cofibrant propositions; and the connection algebra
structure we assume for the interval $\I$ (axioms $\pos{3}$ and
$\pos{4}$ in Figure~\ref{fig:axi}) is weaker than being a de~Morgan
algebra, since we can avoid the use of a de~Morgan involution
operation.  In Section~\ref{sec:relfw} we conclude by considering
other related work.

\paragraph*{\textbf{Agda formalisation}} The definitions and
constructions we carry out in the internal type theory of toposes are
sufficiently involved to warrant machine-assisted formalisation. Our
tool of choice is Agda~\cite{Agda}. We persuaded it to provide an
impredicative universe of mere propositions~\cite[Section~3.3]{HoTT}
using a method due to Escardo~\cite{EscardoM:imp}. This gives an
intensional, proof-relevant version of the subobject classifier
$\Prop$ and of the type theory described in
Section~\ref{sec:intttt}. To this we add postulates corresponding to
the axioms in Figure~\ref{fig:axi}. We also made modest use of the
facility for user-defined rewriting in recent versions of
Agda~\cite{CockxJ:spreyv}, in order to make the connection algebra
axioms $\pos{3}$ and $\pos{4}$ definitional, rather than just
propositional equalities, thereby eliminating a few proofs in favour
of computation.  Using Agda required us to construct and pass around
proof terms that are left implicit in the paper version; we found this
to be quite bearable and also invaluable for getting the details
right. Our development can be found at
\url{https://doi.org/10.17863/CAM.21675}.

\section{Internal Type Theory of a Topos}
\label{sec:intttt}

We rely on the categorical semantics of dependent type theory in terms
of \emph{categories with families} (CwF)~\cite{DybjerP:intt}. For each
topos $\E$ (with subobject classifier $\True:1\fun\Prop$) one can find
a CwF with the same objects, such that the category of families at
each object $X$ is equivalent to the slice category $\E/X$. This can
be done in a number of different ways; for example
\cite[Example~6.14]{PittsAM:catl}, or the more recent references
\cite[Section~1.3]{KapulkinC:simmuf},~\cite{LumsdainePL:locumo} and
\cite{AwodeyS:natmht}, which cater for categories more general than a
topos (and for contextual/comprehension categories rather than CwFs in
the first two cases).  Using the objects, families and elements of
this CwF as a signature, we get an internal type theory along the
lines of those discussed in \cite{MaiettiME:modcdt}, canonically
interpreted in the CwF in the standard
fashion~\cite{HofmannM:synsdt}. We make definitions and postulates in
this internal language for $\E$ using a concrete syntax inspired by
Agda~\cite{Agda}. Dependent function types are written as
$(x:A)\fun B$; their canonical terms are function abstractions,
written as $\lambda(x:A)\fun t$.  Dependent product types are written
as $(x:A)\times B$; their canonical terms are pairs, written as
$(s,t)$.  In the text we use this language informally, similar to the
way that Homotopy Type Theory is presented in~\cite{HoTT}. For
example, the typing contexts of the judgements in the formal version,
such as $[x_0 : A_0, x_1 : A_1(x_0), x_2 : A_2(x_0, x_1)]$, become
part of the running text in phrases like ``given $x_0 : A_0$,
$x_1 : A_1(x_0)$ and $x_2 : A_2(x_0, x_1)$, then\ldots''

In the internal type theory the subobject classifier $\Prop$ of the
topos becomes an impredicative universe of propositions, with logical
connectives ($\True, \False, {\neg}, {\conj}, {\disj}, {\imp}$),
quantifiers ($\forall(x:A), \exists(x:A)$) and equality (${=}$)
satisfying function and proposition extensionality properties.  The
universal property of the subobject classifier gives rise to
comprehension subtypes: given $\Gamma,x:A\ent \varphi(x):\Prop$, then
$\Gamma\ent \{x:A\mid \varphi(x)\}$ is a type whose terms are those
$t:A$ for which $\varphi(t)$ is provable, with the proof being treated
irrelevantly.\footnote{Our Agda development is proof relevant, so that
  terms of comprehension types contain a proof of membership as a
  component.}  Taking $A=1$ to be terminal, for each $\varphi:\Prop$
we have a type whose inhabitation corresponds to provability of~$\varphi$:
\begin{equation}
  \label{eq:extent}
  [\varphi] \defeq \{\_:1 \mid \varphi\}
\end{equation}
We will make extensive use of these types in connection with the
partial elements of a type; see Section~\ref{sec:cofp}.

Instead of quantifying externally over the objects, families and
elements of the CwF associated with $\E$, we will assume $\E$ comes
with an internal full subtopos $\Univ$. In the internal language we
use $\Univ$ as a Russell-style universe (that is, if $A:\Univ$, then
$A$ itself denotes a type) containing $\Prop$ and closed under forming
products, exponentials and comprehension subtypes.

\section{The axioms}
\label{sec:axioms}

In this section we present the axioms that we require to hold in the
internal type theory of a topos $\E$. We provide an overview of each
axiom, giving some intuition as to its purpose and we explain where it
is used in the construction of a model of cubical type theory. This
allows us to see that certain axioms are only required for modelling
specific parts of cubical type theory, for example definitional
identity types (Section~\ref{sec:idet}). These axioms can therefore
be dropped when, for example, looking for models of cubical type
theory with only propositional identity types
(Section~\ref{sec:patt}). For ease of reference the axioms are
collected together in Figure~\ref{fig:axi}, written in the language
described in Section~\ref{sec:intttt}.

\begin{nota}[\textbf{Infix and implicit arguments}]
  In the figure and elsewhere we adopt a couple of useful notational
  conventions from Agda~\cite{Agda}. First, function arguments that
  are written with infix notation are indicated by the placeholder
  notation ``$\_$''; for example $\_\sqcap\_:\I \fun\I \fun \I$
  applied to $i,j:\I$ is written $i \sqcap j$. Secondly, we use the
  convention that braces \{\} indicate implicit arguments; for
  example, the application of $\pos{9}$ in Figure~\ref{fig:axi} to
  $\varphi : \Cof$, $A: [\varphi] \fun \Univ$, $B: \Univ$ and
  $s : (u : [\varphi]) \fun (A\, u \cong B)$ is written
  $\pos{9}\,A\,B\,s$, or $\pos{9}\,\{\varphi\}\,A\,B\,s$ if $\varphi$
  cannot be deduced from the context.
\end{nota}

\begin{figure}
  \centering
  The interval $\I:\Univ$ is connected 
  \[
  \pos{1} :
  [\forall(\varphi:\I\fun\Prop).\;(\forall(i:\I).\;\varphi\,i \disj
  \neg{\varphi\,i}) \imp (\forall(i:\I).\;\varphi\,i) \disj
  (\forall(i:\I).\;\neg\varphi\,i)]
  \]
  has distinct end-points $\src,\tgt:\I$
  \[
  \pos{2} : [\neg(\src=\tgt)]
  \]
  and a connection algebra structure $\_\sqcap\_,\_\sqcup\_:\I
  \fun\I \fun \I$
 \begin{gather*}
   \pos{3} : [\forall(i:\I).\;\src\sqcap x = \src = x\sqcap\src
   \;\conj\; \tgt\sqcap x = x = x\sqcap\tgt]\\
   \pos{4} : [\forall(i:\I).\; \src\sqcup x = x = x \sqcup\src
   \;\conj\; \tgt\sqcup x = \tgt = x\sqcup\tgt].
  \end{gather*}

  \bigskip

  Cofibrant propositions $\Cof=\{\varphi:\Prop\mid \cof\varphi\}$
  (where $\cof:\Prop\fun\Prop$)\\
  include end-point-equality
  \[
  \pos{5} : [\forall(i:\I).\;\cof(i=\src) \;\conj\; \cof(i=\tgt)]
  \]
  and are closed under binary disjunction
  \[
  \pos{6} : [\forall(\varphi\;\psi : \Prop).\; \cof\varphi \imp
  \cof\psi \imp \cof(\varphi \disj \psi)]
  \]
  dependent conjunction
  \[
  \pos{7} : [\forall(\varphi\;\psi : \Prop).\; \cof\varphi \imp
  (\varphi\imp\cof\psi) \imp \cof(\varphi \conj \psi)]
  \]
  and universal quantification over $\I$ 
  \[
  \pos{8} : [\forall(\varphi : \I \fun \Prop).\; (\forall(i : \I).\;
  \cof(\varphi\; i)) \imp \cof(\forall(i : \I).\; \varphi\; i)].
  \]

  \bigskip

  Strictness axiom: any cofibrant-partial type
  $A$\\
  that is isomorphic to a total type $B$ everywhere that $A$ is defined,\\
  can be extended to a total type $B'$ that is isomorphic to $B$:
  \[
  \strictify :
  \begin{array}[t]{@{}l}
    \{ \varphi : \Cof \}
    (A: [\varphi] \fun \Univ)
    (B: \Univ)
    (s : (u : [\varphi]) \fun (A\, u \cong B))
    {}\fun{}\\
    (B' : \Univ) \times \{s' : B' \cong B \mid \forall(u :
    [\varphi]).\; A\, u = B' \conj s\,u = s'\}
  \end{array}
  \]
  \caption{The axioms}
  \label{fig:axi}
\end{figure}

The homotopical approach to type theory~\cite{HoTT} views elements of
identity types as paths between the two elements being equated. We
take this literally, using paths in the topos $\E$ that are morphisms
out of a distinguished object $\I$, called the \emph{interval}. Recall
from Section~\ref{sec:intttt} that we assume the given topos $\E$
comes with a Russell-style universe $\Univ$. We assume that the
interval $\I$ is an element of $\Univ$. We also assume that $\I$ is
equipped with morphisms $\src,\tgt:1\fun \I$ and
$\_\sqcap\_, \_\sqcup\_ : \I \fun \I \fun \I$ satisfying axioms
$\pos{1}$--$\pos{4}$ in Figure~\ref{fig:axi}. The other axioms
($\pos{5}$--$\pos{9}$) concern \emph{cofibrant propositions}, which
are used in Section~\ref{sec:cchmf} to define fibrations, the (indexed
families of) types in the model of cubical type theory.

Axiom~$\pos{1}$ expresses that the interval $\I$ is internally
connected, in the sense that any decidable subset of its elements is
either empty or the whole of $\I$. This implies that if a path in an
inductive datatype has a certain constructor form at one point of the
path, it has the same form at any other point.  This is used at the
end of Section~\ref{sec:comfs} to show that the natural number object
in the topos is fibrant (that is, denotes a type) and that fibrations
are closed under binary coproducts. It also gets used in proving
properties of the glueing construct in
Section~\ref{sec:glueing}. Together with axiom~$\pos{2}$,
connectedness of $\I$ implies that there is no path from $\inl *$ to
$\inr *$ in $1+1$ and hence that the path-based model of Martin-L\"of
type theory determined by the axioms is logically non-degenerate.

Axioms $\pos{3}$ and $\pos{4}$ endow $\I$ with a form of
\emph{connection} algebra structure~\cite{BrownR:douctt}. They
capture some very simple properties of the minimum and maximum
operations on the unit interval $[0,1]$ of real numbers that suffice
to ensure contractibility of singleton types (Section~\ref{sec:patt})
and, in combination with $\pos{2}$, $\pos{5}$ and $\pos{6}$, to define
path lifting from composition for fibrations (see
Section~\ref{sec:comfs}).  In the model of~\cite{CoquandT:cubttc} the
connection algebra structure is given by the lattice structure of the
interval, taking $\_\sqcap\_$ to be binary meet, $\_\sqcup\_$ to be
binary join and using the fact that $\src$ and $\tgt$ are respectively
least and greatest elements.

\begin{rem}[\textbf{De~Morgan involution}]
  \label{rem:demi}
  In the model of \cite{CoquandT:cubttc} $\I$ is not just a lattice,
  but also has an involution operation $1-(\_):\I \fun \I$ (so that
  $(1-(1-i) = i$) making $\sqcup$ the de~Morgan dual of $\sqcap$, in
  the sense that $i \sqcup j = 1-((1 -i) \sqcap(1 -j))$. Although this
  involution structure is convenient, it is not really necessary for
  the constructions that follow. Instead we just give a
  $\src$-version and a $\tgt$-version of certain concepts; for
  example, ``composing from $\tgt$ to $\src$'' as well as ``composing
  from $\src$ to $\tgt$'' in Section~\ref{sec:comfs}.
\end{rem}

Axioms $\pos{2}$--$\pos{6}$ allow us to show that fibrations provide a
model of $\Pi$-{} and $\Sigma$-types; and furthermore to show that the
path types determined by the interval object $\I$
(Section~\ref{sec:patt}) satisfy the rules for identity types
propositionally~\cite{CoquandT:isoe,VanDenBergB:patcpi}.
Axiom~$\pos{7}$ is used to get from these propositional identity types
to the proper, definitional identity types of Martin-L\"of type
theory, via a version of Swan's construction~\cite{SwanA:algwfs}; see
Section~\ref{sec:idet}.

In Section~\ref{univalence} we consider
univalence~\cite[Section~2.10]{HoTT} -- the correspondence between
type-valued paths in a universe and functions that are equivalences
modulo path-based equality. To do so, we first give in
Section~\ref{sec:glueing} a non-strict,
``up-to-isomorphism'' version of the \emph{glueing} construct of
Cohen~\emph{et al} in the internal type theory of the
topos. Axiom~$\pos{8}$ is used in the definition of this weak form of
glueing to ensure that the induced fibration structure extends the
fibration structure on the family that we are
``glueing". Axiom~$\pos{9}$ allows us to regain the strict form of
glueing used by Cohen~\emph{et al}~\cite{CoquandT:cubttc}. Its
validity in presheaf models depends on a construction in the external
meta-theory that cannot be replicated internally; see
Theorem~\ref{thm:strictness} for details.

\section{Path Types}
\label{sec:patt}

Given $A:\Univ$, we call elements of type $\I\fun A$ \emph{paths} in
$A$. The \emph{path type} associated with $A$ is
$\_\path\_: A \fun A \fun \Univ$ where

\begin{equation}
  \label{eq:3}
  a_0 \path a_1 \defeq \{p:\I\fun A \mid p\,\src = a_0 \conj p\,\tgt=
  a_1\} 
\end{equation}
Can these types be used to model the rules for Martin-L\"of identity
types?  We can certainly interpret the identity introduction rule
(reflexivity), since degenerate paths given by constant functions
\begin{equation}
  \label{eq:5}
  \idp a\,i \defeq a
\end{equation}
satisfy ${\idp} : \{A:\Univ\}(a : A) \fun a\path a$. However, we need
further assumptions to interpret the identity elimination rule,
otherwise known as path induction~\cite[Section 1.12.1]{HoTT}. Coquand
has given an alternative (propositionally equivalent) formulation of
identity elimination in terms of substitution functions
$a_0\path a_1 \fun P\,a_0 \fun P\,a_1$ and contractibility of
singleton types $(a_1:A)\times (a_0\path a_1)$;
see~\cite[Figure~2]{CoquandT:modttc}. The connection algebra
structure gives the latter, since using $\pos{3}$ and $\pos{4}$ we
have
\begin{align}
  \kw{ctr} &: \{A:\Univ\}\{a_0\;a_1:A\}(p: a_0\path a_1) \fun (a_0,\idp
  a_0) \path (a_1,p) \label{eq:6}\\
  \kw{ctr}&\,p\,i\defeq (p\,i, \lambda j \fun p(i \sqcap j))\notag
\end{align}
However, to get suitably behaved substitution functions we have to
consider families of types endowed with some extra structure;
and that structure has to lift through the type-forming operations
(products, functions, identity types, etc). This is what the
definitions in the next section achieve.

\section{Cohen-Coquand-Huber-M\"ortberg (CCHM) Fibrations} 
\label{sec:cchmf}

In this section we show how to generalise the notion of fibration
introduced in~\cite[Definition~13]{CoquandT:cubttc} from the
particular presheaf model considered there to any topos with an
interval object as in the previous sections. To do so we use the
notion of \emph{cofibrant proposition} from Figure~\ref{fig:axi} to
internalise the composition and filling operations described in
\cite{CoquandT:cubttc}.

\subsection{Cofibrant propositions}
\label{sec:cofp}

Kan filling and other cofibrancy conditions on collections of
subspaces have to do with extending maps defined on a subspace to maps
defined on the whole space. Here we take ``subspaces of spaces'' to
mean subobjects of objects in toposes. Since subobjects are classified
by morphisms to $\Prop$, it is possible to consider collections of
subobjects that are specified generically by certain propositions.
More specifically, given a property of propositions,
$\cof : \Prop \fun \Prop$, we get a corresponding collection of
propositions
\begin{equation}
  \label{eq:8}
  \Cof \defeq \{ \varphi : \Prop \mid \cof \varphi \} 
\end{equation}
Consider the class of monomorphisms $m:A\mono B$ whose
classifying morphism
\[
\lambda(y:B)\fun \exists(x:A).\;m\,x=y : B\fun\Prop
\]
factors through $\Cof\mono\Prop$. We call such monomorphisms
\emph{cofibrations}. Kan-like filling properties have to do with when a
morphism $A\morphism X$ can be extended along a cofibration
$m:A\mono B$. Instead, working in the internal language of $\E$, we
will consider when partial elements whose domains of definition are in
$\Cof$ can be extended to totally defined elements. Recall that in
intuitionistic logic, partial elements of a type $A$ are often
represented by sub-singletons, that is, by functions $s:A\fun\Prop$
satisfying
\[
\forall(x\;x':A).\; s\,x \conj s\,x' \imp x= x'
\] 
However, it will be more convenient to work with an extensionally
equivalent representation as dependent pairs $\varphi:\Prop$ and
$f:[\varphi]\fun A$, as in the next definition. The proposition
$\varphi$ is the extent of the partial element; in terms of
sub-singletons it is equal to $\exists(x:A).\; s\,x$.

\begin{defi}[\textbf{Cofibrant partial elements, $\Cpf A$}]\label{def:cpe}
  We assume we are given a subobject $\Cof\mono\Prop$ satisfying
  axioms $\pos{5}$--$\pos{8}$ in Figure~\ref{fig:axi}. We call elements
  of type $\Cof$ \emph{cofibrant propositions}.  Given a type
  $A:\Univ$, we define the type of \emph{cofibrant partial elements}
  of $A$ to be
  \begin{equation}
    \label{eq:20}
    \Cpf A \defeq (\varphi:\Cof)\times ([\varphi]\fun A)
  \end{equation}
  An \emph{extension} of such a partial element $(\varphi,f):\Cpf A$
  is an element $a:A$ together with a proof of the following relation: 
  \begin{equation}
    \label{eq:17}
    (\varphi,f)\exto a \defeq \forall(u:[\varphi]).\; f\,u = a
  \end{equation} 
\end{defi}

Note that by taking $i=\src$ in axiom $\pos{5}$ we have
$\cof(\src=\src)$ (that is, $\cof\True$) and $\cof(\src=\tgt)$; and
combining the latter with axiom $\pos{2}$ we deduce also that
$\cof\False$ holds. So $A\mono A$ and $\emptyset\mono A$ are always
cofibrations, where $\emptyset$ is the initial object. Since
$\cof\True$ holds, for every $a : A$ there is a \emph{total} cofibrant
partial element $(\True,\lambda\_\fun a) : \Cpf A$ with $a$ the unique
element that extends $(\True,\lambda\_\fun a)$. Since $\cof\False$
holds, every object $A$ has an empty cofibrant partial element given
by $(\False,\elim_\emptyset) : \Cpf A$ such that every $a : A$ is an
extension of $(\False,\elim_\emptyset)$. (For any $B:\Univ$,
$\elim_\emptyset:[\False]\fun B$ denotes the unique function given by
initiality of $[\False]$.)

\begin{exa}
  \label{exa:1}
  It is helpful to think of variables of type $\I$ as names of
  dimensions in space, so that working in a context $i_1,...,i_n:\I$
  corresponds to working in $n$ dimensions.  Assume that we are
  working in a context with $i,j,k:\I$; this therefore corresponds to
  working in three dimensions.  We think of an element
  $i,j,k:\I \ent a : A$ as a cube in the space $A$, as shown below.
  Let
  $\varphi \defeq (i=\src)\,\disj\,(j=\src)\,\disj\,(j=\tgt\conj
  k=\tgt)$.
  From $\pos{5}$--$\pos{7}$ we have $i,j,k:\I\ent \varphi : \Cof$. We
  think of $\varphi$ as specifying certain faces and edges of a cube,
  in this case the bottom face ($i=\src$), the left face ($j=\src$)
  and the front-right edge ($j=\tgt \conj k=\tgt$), as in the
  right-hand picture below. Then a cofibrant partial element
  $f : [\varphi] \fun A$ can be thought of as a partial cube, only
  defined on the
  region specified by $\varphi$.\\

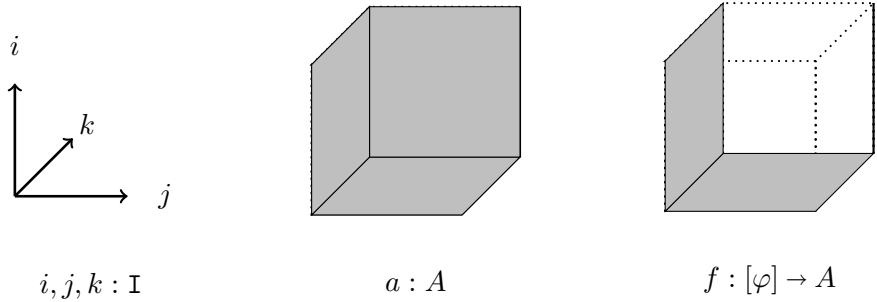
  \begin{minipage}{0.25\linewidth}
    
    \begin{center}
      \vspace{11pt}
      \begin{tikzpicture}
        \draw[->,black,line width=1] (-2,0,0) -- ++(0,1.5,0);
        \node at (-2,2,0) {$i$};
        \draw[->,black,line width=1] (-2,0,0) -- ++(1.5,0,0);
        \node at (0,0,0) {$j$};
        \draw[->,black,line width=1] (-2,0,0) -- ++(0,0,-2);
        \node at (-2,0,-2.5) {$k$};
      \end{tikzpicture}
    \end{center}\
    \[i,j,k:\I\]
  \end{minipage}
  \begin{minipage}{0.3\linewidth}
    
    \begin{center}
      \begin{tikzpicture}
        \draw[dotted,black,line width=0.7] (0,0,0) -- ++(0,2,0) -- cycle;
        \draw[dotted,black,line width=0.7] (0,0,-2) -- ++(0,2,0) -- cycle;
        \draw[dotted,black,line width=0.7] (-2,0,-2) -- ++(0,2,0) -- cycle;
        \draw[dotted,black,line width=0.7] (-2,0,0) -- ++(0,2,0) -- cycle;
\draw[dotted,black,line width=0.7] (0,2,0) -- ++(-2,0,0) -- ++(0,0,-2) --
++(2,0,0) -- cycle;
\draw[black,fill=lightgray] (0,0,0) -- ++(-2,0,0) -- ++(0,0,-2) -- ++(2,0,0) --
cycle;
        
\draw[black,fill=lightgray,line width=0.8] (0,0,-2) -- ++(0,2,0) -- cycle;
\draw[black,fill=lightgray] (-2,0,-2) -- ++(0,0,02) -- ++(0,2,0) -- ++(0,0,-2)
-- cycle;
        
\draw[black,fill=lightgray] (0,0,0) -- ++(-2,0,0) -- ++(0,0,-2) -- ++(2,0,0) --
cycle;
\draw[black,fill=lightgray] (0,0,-2) -- ++(0,2,0) -- ++(-2,0,0) -- ++(0,-2,0)
-- cycle;
\draw[black,fill=lightgray] (-2,0,-2) -- ++(0,0,2) -- ++(0,2,0) -- ++(0,0,-2)
-- cycle;
      \end{tikzpicture}
    \end{center}\
    \[a : A\]
  \end{minipage}
  \begin{minipage}{0.3\linewidth}
    
    \begin{center}
      \begin{tikzpicture}
        \draw[dotted,black,line width=0.7] (0,0,0) -- ++(0,2,0) -- cycle;
        \draw[dotted,black,line width=0.7] (0,0,-2) -- ++(0,2,0) -- cycle;
        \draw[dotted,black,line width=0.7] (-2,0,-2) -- ++(0,2,0) -- cycle;
        \draw[dotted,black,line width=0.7] (-2,0,0) -- ++(0,2,0) -- cycle;
\draw[dotted,black,line width=0.7] (0,2,0) -- ++(-2,0,0) -- ++(0,0,-2) --
++(2,0,0) -- cycle;
\draw[black,fill=lightgray] (0,0,0) -- ++(-2,0,0) -- ++(0,0,-2) -- ++(2,0,0) --
cycle;
        
\draw[black,fill=lightgray,line width=0.8] (0,0,-2) -- ++(0,2,0) -- cycle;
\draw[black,fill=lightgray] (-2,0,-2) -- ++(0,0,02) -- ++(0,2,0) -- ++(0,0,-2)
-- cycle;
      \end{tikzpicture}
    \end{center}\
    \[f : [\varphi] \fun A\]
  \end{minipage}
\end{exa}

\begin{defi}[\textbf{Join of compatible partial elements}]
  \label{def:join}
  Say that two partial elements
  $f:[\varphi]\fun A$ and
  $g:[\psi]\fun A$ are \emph{compatible} if they agree
  wherever they are both defined:
  \begin{equation}
    (\varphi,f)\compat(\psi,g) \defeq \forall(u:[\varphi])(v:[\psi]).\;
    f\,u = g\,v
  \end{equation}
  In that case we can form their join
  $f\, \join\, g: [\varphi\,\disj\,\psi] \fun A$, such that 
  \[
    \forall(u : [\varphi]).\; (f\, \join\, g)\,u = f\, u
    \qquad\qquad
    \forall(v : [\psi]).\; (f\, \join\, g)\,v = g\, v
  \]
  To see why, consider the following pushout square in the topos:
  \[
  \xymatrix@=1pc{*+<1.5em>{[\varphi\conj\psi]} \ar@{>->}[r]
    \ar@{>->}[d] & *+<1.5em>{[\psi]}
    \ar@{>->}[d] \ar@/^1pc/[ddr]^g &\\
    *+<1.5em>{[\varphi]} \ar@/_1pc/[rrd]_f \ar@{>->}[r] & {[\varphi\disj
      \psi]} \ar@{.>}[rd] |-{f\join g}
    \pushoutcorner &\\
    && A}
  \]
  The outer square commutes because
  $(\varphi,f)\compat(\psi,g)$ holds and then $f\join g$ is the unique
  induced morphism out of the pushout. Note that axiom $\pos{6}$ in
  Figure~\ref{fig:axi} implies that the collection of cofibrant
  partial elements is closed under taking binary joins of compatible
  partial elements.
\end{defi}

The following lemma gives an alternative characterization of
axioms~$\pos{7}$ and $\pos{8}$. Since we noted above that $\cof\True$
holds, part (i) of the lemma tells us that cofibrations form a
\emph{dominance} in the sense of synthetic domain
theory~\cite{RosoliniG:conet}; we only use this property of $\Cof$ in
order to construct definitional identity types from propositional
identity types (see Section~\ref{sec:idet}).

\begin{lem}
  \label{lem:ax7-8}
  \begin{itemize}[leftmargin=9mm]
  \item[(i)] Axiom $\pos{7}$ is equivalent to requiring the class of
    cofibrations to be closed under composition.

  \item[(ii)] Axiom $\pos{8}$ is equivalent to requiring the class of
  cofibrations to be closed under exponentiation by $\I$.
  \end{itemize}
  
\end{lem}
\proof
  For part (i), first suppose that $\pos{7}$ holds and that $f:A\mono B$ and
  $g:B\mono C$ are cofibrations. So both
  $\forall(b:B).\;\cof(\exists(a:A).\;f\,a=b)$ and
  $\forall(c:C).\;\cof(\exists(b:B).\;g\,b=c)$ hold and we wish to
  prove $\forall(c:C).\;\cof(\exists(a:A).\;g(f\,a)=c)$. Note that for
  $b:B$ and $c:C$
  \begin{align*}
    g\,b = c &\;\imp\; (\exists(a:A).\;g(f\,a)=c) = (\exists(a:A).\;f\,a=b)
    \qquad\text{since $g$ is a monomorphism}\\
             &\;\imp\; \cof (\exists(a:A).\;g(f\,a)=c) =
               \cof(\exists(a:A).\;f\,a=b)=\True
  \end{align*}
  So for $\varphi\defeq \exists(b:B).\;g\,b=c$ and
  $\psi \defeq \exists(a:A).\;g(f\,a)=c$, we have $\cof\varphi$ and
  $\varphi\imp\cof\psi$. Therefore by $\pos{7}$ we get
  $\cof(\phi\conj \psi)$, which is equal to $\cof(\psi)$ since
  $\psi\imp\varphi$. So we do indeed have
  $\forall(c:C).\;\cof(\exists(a:A).\;g(f\,a)=c)$.
  
  Conversely, suppose cofibrations are closed under composition
  and that $\varphi,\psi : \Prop$ satisfy $\cof\varphi$ and
  $\varphi\imp\cof\psi$. That $\cof\varphi$ holds is equivalent to the
  monomorphism $[\varphi]\mono 1$ being a cofibration; and since 
  \[
  \varphi \;\imp\; (\psi = {\varphi\conj\psi}) \;\imp\; (\cof\psi =
  \cof(\varphi\conj\psi))
  \]
  from $\varphi\imp\cof\psi$ we get $\varphi\imp\cof(\varphi\conj\psi)$
  and hence the monomorphism $[\varphi\conj \psi]\mono[\varphi]$
  is  a cofibration. Composing these monomorphisms, we have that
  $[\varphi\conj\psi]\mono 1$ is a cofibration, that is,
  $\cof(\varphi\conj \psi)$ holds.

  For part~(ii), first suppose that $\pos{8}$ holds and that
  $f:A\mono B$ is a cofibration. We have to show
  that $\I\fun f: (\I\fun A)\mono(\I\fun B)$ is also a
  cofibration. Given $\beta:\I\fun B$ we
  have
  \begin{align*}
    &(\forall i:\I)(\exists a:A).\;f\,a=\beta\,i\\
    &\imp (\forall i:\I)(\exists! a:A).\;f\,a=\beta\,i
    &&\text{(since $f$ is a monomorphism)}\\
    &\imp (\exists \alpha:\I\fun A)(\forall i:\I).\;f(\alpha\,i) =
      \beta\,i
    &&\text{(by unique choice in the topos)}\\
    &\imp (\forall i:\I)(\exists a:A).\;f\,a=\beta\,i  
  \end{align*}
  so that $(\forall i:\I)(\exists a:A).\;f\,a=\beta\,i$ is equal to
  $(\exists \alpha:\I\fun A)(\forall i:\I).\;f(\alpha\,i) = \beta\,i$;
  and the latter is equal to
  $(\exists \alpha:\I\fun A).\; (\I\fun f)\,\alpha=\beta$ by function
  extensionality in the topos. Since $f$ is a cofibration, for each
  $i:\I$ we have $\cof(\exists(a:A).\;f\,a=\beta\,i)$. Hence by
  axiom~$\pos{8}$ we also have
  $\cof((\forall i:\I)(\exists a:A).\;f\,a=\beta\,i)$, that is,
  $\cof(\exists \alpha:\I\fun A).\; (\I\fun f)\,\alpha=\beta)$, as
  required for $\I\fun f$ to be a cofibration.

  Conversely, suppose cofibrations are closed under $\I\fun(\_)$ and
  that $\varphi:\I\fun\Omega$ satisfies $(\forall
  i:\I).\;\cof(\varphi\,i)$. The latter implies that $\{i:\I \mid
  \varphi\,i\} \mono \I$ is a cofibration. Hence so is the
  monomorphism $(\I\fun \{i:\I \mid \varphi\,i\}) \mono (\I\fun\I)$. Since
  $\id:\I\fun\I$ is in the image of this monomorphism iff $(\forall
  i:\I).\; \varphi\,i$ holds, we have $\cof((\forall
  i:\I).\; \varphi\,i)$, as required for axiom $\pos{8}$. \qed

\subsection{Composition and filling structures}
\label{sec:comfs}

Axioms $\pos{5}$--$\pos{7}$ in Figure~\ref{fig:axi} give the simple
properties of cofibrant propositions we use to define an internal
notion of fibration generalising Definition~13
of~\cite{CoquandT:cubttc} and to show that it is closed under forming
$\Sigma$-, $\Pi$- and $\Id$-types, as well as basic datatypes.

Given an interval-indexed family of types $A:\I\fun\Univ$, we think of
elements of the dependent function type
$\Pi_\I A \defeq (i:\I)\fun A\,i$ as dependently typed paths. We call
elements of type $\Cpf(\Pi_\I A)$ \emph{cofibrant-partial
  paths}. Given $(\varphi,f): \Cpf(\Pi_\I A)$, we can evaluate it at a
point $i:\I$ of the interval to get a cofibrant partial element
$(\varphi,f)\conc i : \Cpf(A\,i)$:
\begin{equation}
  \label{eq:22}
  (\varphi,f)\conc i \defeq (\varphi, \lambda(u:[\varphi])\fun
  f\,u\,i) 
\end{equation}
An operation for \emph{filling from $\src$} in $A:\I\fun\Univ$ takes
any $(\varphi,f): \Cpf(\Pi_\I A)$ and any $a_0:A\,\src$ with
$(\varphi,f)\conc\src \exto a_0$ and extends $(\varphi,f)$ to a
dependently typed path $g:\Pi_\I A$ with $g\,\src=a_0$. This is a form
of uniform \emph{Homotopy Extension and Lifting Property}
(HELP)~\cite[Chapter~10, Section~3]{MayPJ:concat} stated internally in
terms of cofibrant propositions rather than externally in terms of
cofibrations. A feature of our internal approach compared with
Cohen~\emph{et al} is that their \emph{uniformity} condition on
composition/filling operations~\cite[Definition~13]{CoquandT:cubttc},
which allows one to avoid the non-constructive aspects of the
classical notion of Kan filling~\cite{CoquandT:krimss}, becomes
automatic when the operations are formulated in terms of the internal
collection $\Cof$ of cofibrant propositions.

\begin{exa}
  For some intuition as to why such an operation is referred to as
  filling, consider the following example. For simplicity, assume that
  $A : \I\fun\Univ$ is a constant family
  $A \defeq \lambda(\_:\I) \fun A'$.  Recall that we think of
  variables of type $\I$ as dimensions in space; so that, given an
  element $a : A$ in an ambient context $j,k:\I$, we think of $a$ as a
  square in the space $A$. We are interested in extending this two
  dimensional square to a three dimensional cube as indicated below.
  \begin{center}
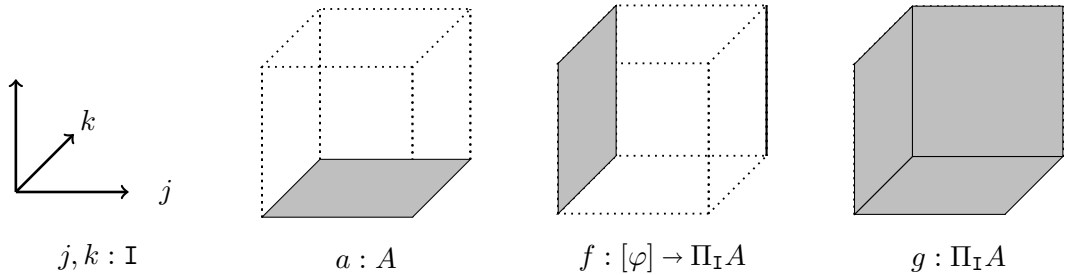
    \begin{minipage}{0.2\linewidth}
      \begin{center}
        \vspace{27pt}
        \begin{tikzpicture}
          \draw[->,black,line width=1] (-2,0,0) -- ++(0,1.5,0);
          \draw[->,black,line width=1] (-2,0,0) -- ++(1.5,0,0);
          \node at (0,0,0) {$j$};
          \draw[->,black,line width=1] (-2,0,0) -- ++(0,0,-2);
          \node at (-2,0,-2.5) {$k$};
        \end{tikzpicture}
      \end{center}
      \[
      j,k:\I
      \]
    \end{minipage}
    \begin{minipage}{0.25\linewidth}
      \begin{center}
        \begin{tikzpicture}
          \draw[dotted,black,line width=0.7] (0,0,0) -- ++(0,2,0) -- cycle;
          \draw[dotted,black,line width=0.7] (0,0,-2) -- ++(0,2,0) -- cycle;
          \draw[dotted,black,line width=0.7] (-2,0,-2) -- ++(0,2,0) -- cycle;
          \draw[dotted,black,line width=0.7] (-2,0,0) -- ++(0,2,0) -- cycle;
\draw[dotted,black,line width=0.7] (0,2,0) -- ++(-2,0,0) -- ++(0,0,-2) --
++(2,0,0) -- cycle;
\draw[black,fill=lightgray] (0,0,0) -- ++(-2,0,0) -- ++(0,0,-2) -- ++(2,0,0) --
cycle;
        \end{tikzpicture}
      \end{center}
      \[
      a : A
      \]
    \end{minipage}
    \begin{minipage}{0.25\linewidth}
      \begin{center}
        \begin{tikzpicture}
\draw[dotted,black, line width=0.7] (0,0,0) -- ++(-2,0,0) -- ++(0,0,-2) --
++(2,0,0) -- cycle;
          
          \draw[dotted,black,line width=0.7] (0,0,0) -- ++(0,2,0) -- cycle;
          \draw[dotted,black,line width=0.7] (0,0,-2) -- ++(0,2,0) -- cycle;
          \draw[dotted,black,line width=0.7] (-2,0,-2) -- ++(0,2,0) -- cycle;
          \draw[dotted,black,line width=0.7] (-2,0,0) -- ++(0,2,0) -- cycle;
\draw[dotted,black,line width=0.7] (0,2,0) -- ++(-2,0,0) -- ++(0,0,-2) --
++(2,0,0) -- cycle;
          
          \draw[black,fill=gray,line width=0.8] (0,0,-2) -- ++(0,2,0) -- cycle;
\draw[black,fill=lightgray] (-2,0,-2) -- ++(0,0,02) -- ++(0,2,0) -- ++(0,0,-2)
-- cycle;
        \end{tikzpicture}
      \end{center}
      \[
      f : [\varphi] \fun \Pi_\I A
      \]
    \end{minipage}
    \begin{minipage}{0.25\linewidth}
      \begin{center}
        \begin{tikzpicture}
          
          \draw[dotted,black,line width=0.7] (0,0,0) -- ++(0,2,0) -- cycle;
          \draw[dotted,black,line width=0.7] (0,0,-2) -- ++(0,2,0) -- cycle;
          \draw[dotted,black,line width=0.7] (-2,0,-2) -- ++(0,2,0) -- cycle;
          \draw[dotted,black,line width=0.7] (-2,0,0) -- ++(0,2,0) -- cycle;
\draw[dotted,black,line width=0.7] (0,2,0) -- ++(-2,0,0) -- ++(0,0,-2) --
++(2,0,0) -- cycle;
\draw[black,fill=lightgray] (0,0,0) -- ++(-2,0,0) -- ++(0,0,-2) -- ++(2,0,0) --
cycle;
\draw[black,fill=lightgray] (0,0,-2) -- ++(0,2,0) -- ++(-2,0,0) -- ++(0,-2,0)
-- cycle;
\draw[black,fill=lightgray] (-2,0,-2) -- ++(0,0,02) -- ++(0,2,0) -- ++(0,0,-2)
-- cycle;
        \end{tikzpicture}
      \end{center}
      \[
      g : \Pi_\I A 
      \]
    \end{minipage}
  \end{center}
  However, let us imagine that we already know how to extend $a$ on
  certain faces and edges of the cube, for example, on the faces/edges
  specified by $\varphi \defeq (j=\src)\,\disj\,(j=\tgt\conj k=\tgt)$.
  This means that we have a cofibrant partial path
  $f : [\varphi] \fun \Pi_\I A$ which agrees with $a$ where they are
  both defined, that is $(\varphi,f)\conc \src \exto a$. Note that $f$
  is a partial path rather than partial element because, on the
  faces/edges where it is defined, it must be defined at all points
  along the new dimension by which we are extending $a$,
  i.e. $\varphi$ cannot depend on this new dimension. A filling for
  this data is a cube $g : \Pi_\I A$ which agrees with the faces/edges
  that we started with. That is, it extends $f$ and agrees with $a$ at
  the base of the cube: $(\varphi,f)\exto g$ and $g\,\src=a$.
\end{exa}

Since we are not assuming any structure on the interval for reversing
paths (see Remark~\ref{rem:demi}), we also need to consider the
symmetric notion of filling from $\tgt$. Let
\begin{equation}
  \label{eq:12}
  \Bool \defeq\{i:\I \mid {i=\src}\disj{i=\tgt}\}
\end{equation}
Note that because of axiom
$\pos{2}$, this is isomorphic to the object of Booleans,
$1+1$ and hence there is a function
\begin{equation}
  \label{eq:15}
  \inv{\rule{0pt}{.7em}\;\;} : \Bool \fun \Bool
\end{equation}
satisfying $\inv{\src}=\tgt$ and $\inv{\tgt}=\src$. In what follows,
instead of using path reversal we parameterise definitions with
$e:\Bool$ and use \eqref{eq:15} to interchange $\src$ and $\tgt$.

\begin{defi}[\textbf{Filling structures}]
  
  Given $e:\Bool$, the type $\Fill e\,A :\Univ$ of \emph{filling
    structures} for an $\I$-indexed families of types $A:\I\fun\Univ$,
  is defined by:
  \begin{equation}
    \label{eq:7}
    \Fill e\, A \defeq 
    \begin{array}[t]{@{}l}
      (\varphi:\Cof)(f: [\varphi]\fun\Pi_\I A)(a: \{a' : A\,e \mid
      (\varphi,f)\conc e \exto a'\}) \fun {}\\
      \{g: \Pi_\I A \mid
      (\varphi,f)\exto g \;\conj\; g\,e=a\}   
    \end{array}
  \end{equation}
\end{defi}

A notable feature of \cite{CoquandT:cubttc} compared with preceding
work~\cite{CoquandT:modttc} is that such filling structure can be
constructed from a simpler \emph{composition} structure that just
produces an extension at one end of a cofibrant-partial path from an
extension at the other end. We will deduce this using axioms
$\pos{3}$--$\pos{6}$ from the following, which is the main notion of
this paper.

\begin{defi}[\textbf{CCHM fibrations}]%
\label{def:cchm-fib}
  A \emph{CCHM fibration} $(A,\alpha)$ over a type $\Gamma:\Univ$ is a
  family $A:\Gamma\fun \Univ$ equipped with a fibration structure
  $\alpha : \isFib A$, where
  $\isFib : \{\Gamma:\Univ\}(A: \Gamma\fun\Univ) \fun \Univ$ is
  defined by
  \begin{equation}
    \label{eq:39}
    \isFib\,\{\Gamma\}\,A \defeq (e:\Bool)(p:\I\fun \Gamma) \fun \Comp
    e\, (A\comp p) 
  \end{equation}
  Here ${\Comp} : (e:\Bool)(A:\I\fun\Univ) \fun\Univ$ is the type
  of \emph{composition structures} for $\I$-indexed families:
  \begin{equation}
    \label{eq:23}
    \Comp e\, A \defeq 
    \begin{array}[t]{@{}l}
      (\varphi:\Cof)(f: [\varphi]\fun\Pi_\I A)\fun{}\\
      \{a_0:A\,e \mid (\varphi,f)\conc e \exto a_0\} \fun
      \{a_1: A\,\inv{e} \mid (\varphi,f)\conc\inv{e} \exto a_1\}  
    \end{array}
  \end{equation}
  Unwinding the definition, if $\alpha: \isFib A$ then $\alpha\,\src$
  satisfies that for each cofibrant partial path
  $f: [\varphi]\fun\Pi_\I (A \comp p)$ over a path $p:\I\fun\Gamma$,
  if $a_0:A\,\src$ extends the partial element $(\varphi,f)\conc\src$,
  \emph{i.e.}~$\forall(u:[\varphi]).\; f\,u\,\src = a_0$, then
  $\alpha\,\src\,p\,\varphi\,f\,a_0 : A\,\tgt$ extends
  $(\varphi,f)\conc\tgt$,
  \emph{i.e.}~$\forall(u:[\varphi]).\; f\,u\,\tgt =
  \alpha\,\src\,p\,\varphi\,f\,a_0$; and similarly for $\alpha\,\tgt$.
\end{defi}

\begin{defi}[\textbf{The CwF of CCHM fibrations}]%
\label{def:cchm-fib-cwf}
  Let $\Fib{\Gamma}$ be the type of CCHM fibrations over an object
  $\Gamma$, defined by
  \begin{equation}
    \Fib\,\Gamma \defeq (A : \Gamma \to \Univ) \times \isFib{A}
  \end{equation}
  CCHM fibrations are closed under re-indexing: given
  $\gamma:\Delta\fun\Gamma$ and $A:\Gamma\fun\Univ$, we get a function
  $\_[\gamma]: \isFib A \fun \isFib(A\comp\gamma)$ defined by
  $\alpha[\gamma]\,e\,p \defeq \alpha\,e\,(\gamma\comp p)$. 
  Therefore we get a function
  $\_[\_] : (\Delta \fun \Gamma) \fun \Fib{\Gamma} \fun \Fib{\Delta}$
  given by
  \begin{equation}
    \label{eq:21}
    (A,\alpha)[\gamma] \defeq (A \circ \gamma, \alpha[\gamma])
  \end{equation}
  which is functorial: $((A,\alpha)[\id] = (A,\alpha)$ and
  $(A,\alpha)[g\comp f] = (A,\alpha)[g][f]$. It follows that $\Fib$
  has the structure of a Category with Families by taking families to
  be CCHM fibrations $(A,\alpha) : \Fib{\Gamma}$ over each
  $\Gamma:\Univ$ and elements of such a family to be dependent
  functions in $(x:\Gamma)\fun A\,x$.
\end{defi}

\begin{rem}[\textbf{Fibrant objects}]%
  \label{rem:fibo}
  We say $A:\Univ$ is a \emph{fibrant object} if we have a fibration
  structure for the constant family $\lambda(\_:1)\fun A$ over the
  terminal object $1$.  Note that if $(A,\alpha):\Fib{\Gamma}$ is a
  fibration, then for each $x:\Gamma$ the type $A\,x:\Univ$
  is fibrant, with the fibration structure given by reindexing $\alpha$
  by the map $\lambda(\_:1)\fun x : 1 \fun \Gamma$.
  However the converse is not true: having a family of
  fibration structures, that is, an element of
  $(x:\Gamma)\fun \isFib(\lambda(\_:1)\fun A\,x)$, is weaker than
  having a fibration structure for $A:\Gamma\fun\Univ$. To see why,
  consider the family, $P : \I \fun \Univ$ defined by
  \begin{equation}
    P\; i \defeq [i = \src]
  \end{equation}
  For each $i:\I$ the fibre $P\, i : \Univ$ is a fibrant object, with
  a fibration structure, $\rho_i : \isFib(\lambda(\_:1)\fun P\,i)$,
  given by $\rho_i\;e\;p\;\phi\;f\;x \defeq x$.  However, it is not
  possible to construct a $\rho : \isFib{P}$. For if it were, then we
  could define
  $\rho\;\src\;\id\;\False\;\elim_\emptyset\;* : [1 = 0]$; combined
  with $\pos{2}$, this would lead to contradiction.
\end{rem}

If $\alpha:\Fill e\,A$, then
$\lambda \varphi\,f\,a \fun \alpha\,\varphi\,f\,a\,\inv{e} : \Comp e\,
A$
and so every filling structure gives rise to a composition structure.
Conversely, the composition structure of a CCHM fibration gives rise
to filling structure:

\begin{lem}[\textbf{Filling structure from composition structure}]
  \label{lem:fill}
  Given $\Gamma:\Univ$, $A:\Gamma\fun\Univ$, $e:\Bool$,
  $\alpha:\isFib A$ and $p:\I\fun \Gamma$, there is a filling
  structure $\fil e\,\alpha\,p : \Fill e\,(A\comp p)$ that agrees with
  $\alpha$ at $\inv{e}$, that is:
  \begin{multline}
    \label{eq:1}
    \forall(\varphi:\Cof)(f:[\varphi]\fun\Pi_\I A)(a : A(p\,e)).\\
    (\varphi,f)\conc e \exto a \;\imp\;
    \fil e\,\alpha\,p\,\varphi\,f\,a\,\inv{e} =
    \alpha\,e\,p\,\varphi\,f\,a
  \end{multline}
  Furthermore, $\fil$ is stable under re-indexing in the sense that
  for all $\gamma:\Delta\fun\Gamma$ and $p:\I\fun\Delta$
  \begin{equation}
    \label{eq:18}
    \fil e\,\alpha\,(\gamma\comp p) = \fil
    e\,(\alpha[\gamma])\,p 
  \end{equation}
\end{lem}
\proof 
The construction of filling from composition
follows~\cite[Section~4.4]{CoquandT:cubttc}, but just using the
connection algebra structure on $\I$ (axioms $\pos{3}$ and
$\pos{4}$), rather than a de~Morgan algebra structure. Suppose
$\Gamma :\Univ$, $A :\Gamma\fun\Univ$, $e :\Bool$, $\alpha :\isFib A$,
$p :\I\fun \Gamma$, $\varphi :\Cof$,
$f :[\varphi]\fun\Pi_\I(A\comp p)$, $a : A(p\,e)$ with
$(\varphi,f)\conc e \exto a$, and $i :\I$. Then using
Definition~\ref{def:join} we can define
\begin{equation}
  \fil e\,\alpha\,p\,\varphi\,f\,a\,i \defeq \alpha\, e\,(p'\,i)\,(\varphi
  \disj {i=e})\,(f'\,i\join g\,i)\,a\label{eq:9}
\end{equation}
where
\begin{align*}
  &p':\I\fun \I\fun \Gamma \text{\ is defined by\ }
  p'\,i\,j\defeq p(i \sqcap_e j)\\
  &f' : (i:\I) \fun [\varphi] \fun \Pi_\I(A\comp (p'\,i)) \text{\ is
    defined by\ }
  f'\,i\,u\,j \defeq f\,u\,(i \sqcap_e j)\\
  &g : (i:\I) \fun \{g': [i=e] \fun \Pi_\I(A\comp (p'i)) \mid
  (\varphi,f'\,i)\compat(i=e,g')\} \text{\ is defined by\ } g\,i\,v\,j
  \defeq a
\end{align*}
and where $\sqcap_e$ is given by ${\sqcap_{\src}}\defeq {\sqcap}$
and ${\sqcap_{\tgt}}\defeq {\sqcup}$. Finally, property \eqref{eq:18}
is immediate from definitions \eqref{eq:21} and \eqref{eq:9}.
\qed

Compared with~\cite{CoquandT:modttc}, the fact that filling can be
defined from composition considerably simplifies the process of
lifting fibration structure through the usual type-forming constructs,
as the following two theorems demonstrate. Their proofs are
internalisations of those in~\cite[Section~4.5]{CoquandT:cubttc},
except that we avoid the use Cohen~\emph{et al} make of de~Morgan
involution.

\begin{thm}[\textbf{Fibrant $\Sigma$-types}]
  \label{thm:fib-sigma}
  There is a function
  \begin{equation}
    {\isFib_\Sigma} :
    \begin{array}[t]{@{}l}
      \{\Gamma:\Univ\}\{A_1:\Gamma\fun\Univ\}\{A_2:(x:\Gamma)\times
      A_1\,x\fun\Univ\} \fun{}\\ 
      \isFib A_1 \fun \isFib A_2 \fun \isFib (\Sigma\,A_1\,A_2)
    \end{array}\label{eq:4}
  \end{equation}
  where $\Sigma\,A_1\,A_2\,x \defeq (a_1: A_1\,x) \times
  A_2(x,a_1)$.
  The function is stable under re-indexing, in the sense that for all
  $\gamma:\Delta\fun\Gamma$
  \begin{equation}
    \label{eq:16}
    (\isFib_\Sigma\alpha_1\,\alpha_2)[\gamma] =
    \isFib_\Sigma(\alpha_1 [\gamma])(\alpha_2[\gamma\times\id]) 
  \end{equation}
  Hence the category with families given by CCHM fibrations supports
  the interpretation of
  $\Sigma$-types~\cite[Definition~3.18]{HofmannM:synsdt}.
\end{thm}
\proof The construction of $\isFib_\Sigma$ makes use of the filling
operation from Lemma~\ref{lem:fill}.  Given $\Gamma :\Univ$,
$A_1 :\Gamma\fun\Univ$, $A_2 :(x:\Gamma)\times A_1\,x\fun\Univ$,
$\alpha_1 :\isFib A_1$, $\alpha_2 :\isFib A_2$, $e :\Bool$,
$p :\I\fun \Gamma$, $\varphi :\Cof$,
$f :[\varphi]\fun \Pi_\I((\Sigma\,A_1\,A_2)\comp p)$ and
$(a_1,a_2) :(\Sigma\,A_1\,A_2)(p\,e)$ with
$(\varphi,f)\conc e \exto (a_1,a_2)$, define
\begin{equation}
  \isFib_\Sigma\,\alpha_1\,\alpha_2\,e\,p\,\varphi\,f\,(a_1,a_2)
  \defeq (\alpha_1\,e\,p\,\varphi\,f_1\,a_1 \;,\;
  \alpha_2\,e\,q\,\varphi\,f_2\,a_2)\label{eq:24}
\end{equation}
where
\begin{align*}
  &f_1 : [\varphi] \fun \Pi_\I(A_1\comp p)\\
  &f_1\,u\,i \defeq \fst(f\,u\,i)\\
  &q : \I \fun (x:\Gamma)\times A_1\,x\\
  &q \defeq
    \Pair{p}{\fil e\,\alpha_1\,p\,\varphi\,f_1\,a_1}\\
  &f_2 : [\varphi] \fun \Pi_\I(A_2\comp q)\notag\\
  &f_2\,u\,i \defeq \snd(f\,u\,i)\notag
\end{align*}
Thus
$\isFib_\Sigma\,\alpha_1\,\alpha_2\,e\,p\,\varphi\,f\,(a_1,a_2) :
(\Sigma\,A_1\,A_2)(p\,\inv{e})$; and since
\begin{align*}
  &\forall(u:[\varphi]).\; f_1\,u\,\inv{e} =
    \alpha_1\,e\,p\,\varphi\,f_1\,a_1 =
    \fil e\,\alpha_1\,p\,\varphi\,f_1\,a_1\,\inv{e}\\ 
  &\forall(u:[\varphi]).\;f_2\,u\,\inv{e} =
    \alpha_2\,e\,q\,\varphi\,f_2\,a_2
\end{align*}
hold, it follows that
\[
(\varphi,f)\conc\inv{e} \exto
\isFib_\Sigma\,\alpha_1\,\alpha_2\,e\,p\,\varphi\,f\,(a_1,a_2).
\]
Hence
$\isFib_\Sigma\,\alpha_1\,\alpha_2: \isFib (\Sigma\,A_1\,A_2)$.
Finally, property \eqref{eq:16} follows from \eqref{eq:18} and
\eqref{eq:24}.  
\qed

\begin{thm}[\textbf{Fibrant $\Pi$-types}]
  \label{thm:fib-pi}
  There is a function
  \begin{equation} 
    {\isFib_\Pi}: 
    \begin{array}[t]{@{}l}
      \{\Gamma:\Univ\}\{A_1:\Gamma\fun\Univ\}\{A_2:(x:\Gamma)\times
      A_1\,x\fun\Univ\} \fun{}\\ 
      {\isFib A_1} \fun {\isFib A_2} \fun {\isFib (\Pi\,A_1\,A_2)}
    \end{array}\label{eq:27}
  \end{equation}
  where $\Pi\,A_1\,A_2\,x \defeq (a_1: A_1\,x) \fun A_2(x,a_1)$. This
  function is stable under re-indexing (cf.~\ref{eq:16}) and
  hence the category with families given by CCHM fibrations supports
  the interpretation of
  $\Pi$-types~\cite[Definition~3.15]{HofmannM:synsdt}.
\end{thm}
\proof
Given
$\Gamma :\Univ$, $A_1 :\Gamma\fun\Univ$,
$A_2 :(x:\Gamma)\times A_1\,x\fun\Univ$, $\alpha_1 :\isFib A_1$,
$\alpha_2 :\isFib A_2$, $e :\Bool$, $p :\I\fun \Gamma$,
$\varphi :\Cof$, $f :[\varphi]\fun \Pi_\I((\Pi\,A_1\,A_2)\comp p)$,
$g :(\Pi\,A_1\,A_2)(p\,e)$ with $(\varphi,f)\conc e \exto g$ and
$a_1 :A_1(p\,\inv{e})$, using Lemma~\ref{lem:fill} 
we define
\begin{equation}
  \isFib_\Pi\,\alpha_1\,\alpha_2\,e\,p\,\varphi\,f\,g\,a_1\defeq
  \alpha_2\,e\,q\,\varphi\,f_2\,a_2\label{eq:53}
\end{equation}
where
\begin{align*}
  &f_1 : \Pi_\I (A_1\comp p)\\
  &f_1 \defeq \fil \inv{e}\,\alpha_1\,p\,\False\,\elim_\emptyset\,a_1\\
  &q:\I \notag \fun (x:\Gamma)\times A_1\,x\\
  &q \defeq \Pair{p}{f_1}\\
  &f_2 : [\varphi] \fun \Pi_\I(A_2\comp q)\\
  &f_2\,u\,i \defeq f\,u\,i\,(f_1\,i)\\
  &a_2: \{a_2': A_2(q\,e) \mid (\varphi,f_2)\conc e \exto
    a_2'\}\\ 
  &a_2 \defeq g(f_1\,e)\notag 
\end{align*}
Since we know that
$\fil\inv{e}\,\alpha_1\,p\,\False\,\elim_\emptyset\,a_1\,\inv{e} =
a_1$, therefore we have
\begin{equation}
  \label{eq:32}
  \isFib_\Pi\,\alpha_1\,\alpha_2\,e\,p\,\varphi\,f\,g\,a_1:
  A_2(q\,\inv{e}) = A_2(p\,\inv{e}, f_1\,\inv{e}) =
  A_2(p\,\inv{e}, a_1)
\end{equation}
Furthermore, since
$(\varphi,f_2)\conc\inv{e}\exto \alpha_2\,e\,q\,\varphi\,f_2\,a_2$,
for any $u:[\varphi]$ we have
\begin{equation}
  \label{eq:25}
  f\,u\,\inv{e}\,a_1 = f\,u\,\inv{e}\,(f_1\,\inv{e}) =
  f_2\,u\,\inv{e} = \alpha_2\,e\,q\,\varphi\,f_2\,a_2 =
  \isFib_\Pi\,\alpha_1\,\alpha_2\,e\,p\,\varphi\,f\,g\,a_1 
\end{equation}
Since \eqref{eq:32} and \eqref{eq:25} hold for all
$a_1:A_1(p\,\inv{e})$, from the first if follows that
\[
\isFib_\Pi\,\alpha_1\,\alpha_2\,e\,p\,\varphi\,f\,g :
(\Pi\,A_1\,A_2)(p\,\inv{e})
\]
and from the second that
$(\varphi,f)\conc\inv{e}\exto
\isFib_\Pi\,\alpha_1\,\alpha_2\,e\,p\,\varphi\,f\,g$.
Therefore we have that \eqref{eq:53} does give an element of
$\isFib (\Pi\,A_1\,A_2)$.  Finally, stability of
$\isFib_\Pi\,\alpha_1\,\alpha_2$ under re-indexing follows from
\eqref{eq:18}.  
\qed
  
These theorems allow us to construct fibration structures for
$\Sigma$-{} and $\Pi$-types, given fibration structures for their
constituent types. But are there any fibration structures to begin
with? We answer this question by showing that the natural number
object $\Nat$ in the topos is always fibrant. This is proved for the
topos of cubical sets $\hat{\dm}$ in
\cite[Section~4.5]{CoquandT:modttc} by defining a composition
structure by primitive recursion. We give a more elementary proof
using the fact that the interval object in $\hat{\dm}$ satisfies axiom
$\pos{1}$ (see Theorem~\ref{thm:sat-pos1}).
  
\begin{thm}[\textbf{$\Nat$ is fibrant}]
  \label{thm:fib-nat}
  If $\Nat$ is an object with decidable equality, then there is a
  function
  ${\isFib_{\Nat}} : \{\Gamma:\Univ\} \fun
  \isFib(\lambda(\_:\Gamma)\fun \Nat)$.
  In particular, if the topos $\E$ has a natural number object
  $1\xrightarrow{\Zero}\Nat\xrightarrow{\Succ}\Nat$, then the category
  with families given by CCHM fibrations has a natural number object.
\end{thm}
\proof
  Suppose $\Gamma :\Univ$, $e :\Bool$, $p :\I\fun\Gamma$,
  $\varphi :\Cof$, $f :[\varphi]\fun\Pi_\I(\lambda\_\fun\Nat)$ and
  $n :\Nat$ with $(\varphi,f)\conc e \exto n$.  By assumption on
  $\Nat$, for each $u:[\varphi]$ the property
  $\lambda(i:\I)\fun(f\,u\,i=n) :\I\fun\Prop$ is decidable; hence by
  axiom $\pos{1}$ and the fact that $f\,u\,e=n$, we also have
  $f\,u\,\inv{e} = n$. Therefore we can get
  $\isFib_{\Nat}e\,p\,\varphi\,f\,n :\{n':\Nat \mid
  (\varphi,f)\conc\inv{e}\exto n'\}$
  just by defining: $\isFib_{\Nat}e\,p\,\varphi\,f\,n \defeq n$. For
  the last part of the theorem we use the fact that in a topos with
  natural number object, equality of numbers is decidable.
\qed

A similar use of axiom $\pos{1}$ suffices to prove:

\begin{thm}[\textbf{Fibrant coproducts}]
  \label{thm:fib-coprod}
  Writing $A_1\xrightarrow{\inl}A_1+A_2\xleftarrow{\inr}A_2$ for the
  coproduct of $A_1$ and $A_2$ in $\E$, we lift this to families of
  types,
  $\_\uplus\_ :
  \{\Gamma:\Univ\}(A_1\;A_2:\Gamma\fun\Univ)\fun\Gamma\fun\Univ$,
  by defining $(A_1 \uplus A_2)\,x \defeq A_1\,x + A_2\,x$. Then there
  is a function
  \begin{equation}
    \label{eq:10}
    {\isFib_{\uplus}} :\{\Gamma:\Univ\}\{A_1\; A_2:\Gamma\fun\Univ\} \fun
    \isFib A_1 \fun \isFib A_2 \fun \isFib(A_1\uplus A_2) 
  \end{equation}
  and this fibration structure on coproducts is stable under
  re-indexing. Hence the category with families given by CCHM
  fibrations has binary coproducts.
\end{thm}
\proof The proof makes use of the principle of unique choice, which
holds in the internal type theory of a topos:
\begin{equation}
  \label{eq:31}
  \kw{uc}:(A:\Univ)(\varphi:A\fun\Prop)\fun
  [\exists!(a:A).\;\varphi\,a] \fun \{a:A \mid \varphi\,a\}
\end{equation}
where $\exists!(a:A).\;\varphi\,a \defeq \exists(a:A).\;\varphi\,a
\conj \forall(a':A).\; \varphi\,a' \imp a=a'$.

Suppose we have $\Gamma:\Univ$,
$A_1\; A_2 : \Gamma\fun \Univ$, $\alpha_1 : \isFib A_1$,
$\alpha_2 : \isFib A_2$, $e :\Bool$, $p : \I \fun \Gamma$,
$\varphi : \Cof$, $g : [\varphi] \fun \Pi_\I((A_1\uplus A_2)\comp p)$
and $c : A_1(p\,e) + A_2(p\,e)$ with $(\varphi,g)\conc e \exto c$.
Note that for all $u :[\varphi]$ and $i:\I$
\begin{align*}
  &P_1, P_2 : [\varphi] \fun \I \fun \Prop\\
  &P_1\,u\,i \defeq \exists!(a_1:A_1(p\,i)).\; g\,u\,i =\inl a_1\\
  &P_2\,u\,i \defeq \exists!(a_2:A_2(p\,i)).\; g\,u\,i =\inr a_2 
\end{align*}
are complementary propositions
($P_1\,u\,i \;\conj\; P_2\,u\,i =\False$ and
$P_1\,u\,i \;\disj\; P_2\,u\,i =\True$); and hence by $\pos{1}$ we
have that
$(\forall(i:\I).\;P_1\,u\,i) \disj (\forall(i:\I).\;P_2\,u\,i)$.
Either $c=\inl a_1$ for some $a_1:A_1(p\,e)$, or $c=\inr a_2$ for some
$a_2:A_2(p\,e)$. In the first case, since
$\forall(u:[\varphi]).\; g\,u\,e = c$, it follows that
$\forall(u:[\varphi])(i:\I).\; P_1\,u\,i$; then using $\kw{uc}$ we get
some $f_1: [\varphi]\fun \Pi_\I(A_1\comp p)$ with
$\forall(u:[\varphi])(i:\I).\; g\,u\,i= \inl(f_1\,u\,i)$ and we can
define
\[
\isFib_{\uplus}\alpha_1\,\alpha_2\,e\,p\,\varphi\,g\,(\inl a_1) \defeq
\inl(\alpha_1 e\,p\,\varphi\,f_1\,a_1)
\]
Similarly if $c=\inr a_2$, then there is some
$f_2: [\varphi]\fun \Pi_\I(A_2\comp p)$ with
$\forall(u:[\varphi])(i:\I).\; g\,u\,i= \inr(f_2\,u\,i)$ and we can
define
\[
\isFib_{\uplus}\alpha_1\,\alpha_2\,e\,p\,\varphi\,g\,(\inr a_2) \defeq
\inr(\alpha_2 e\,p\,\varphi\,f_2\,a_2).
\]
\qed

\subsection{Identity types}
\label{sec:idet}

\begin{thm}[\textbf{Fibrant path types}]
  \label{thm:fib-path}
  There is a function
  \begin{equation}
    \label{eq:26}
    {\isFib_{\Path}} : \{\Gamma:\Univ\}\{A:\Gamma\fun\Univ\} \fun
    \isFib A \fun \isFib (\Path A)
  \end{equation}
  where $\Path A : (x:\Gamma)\times (A\,x\times A\,x) \fun \Univ$ is
  given by
  \begin{equation}
    \label{eq:11}
    \Path A\,(x,(a_0,a_1)) \defeq a_0 \path a_1 
  \end{equation}
  and where $\path$ is as in~\eqref{eq:3}.  This fibration structure on
  path types is stable under re-indexing, in the sense that for all
  $\gamma:\Delta\fun\Gamma$
  \begin{equation}
    \label{eq:28}
    (\isFib_{\Path}\alpha)[\gamma\times(\id\times\id)] =
    \isFib_{\Path}(\alpha[\gamma])
  \end{equation}
\end{thm}
\proof Given $\Gamma :\Univ$, $A :\Gamma\fun\Univ$,
$\alpha :\isFib A$, $e :\Bool$,
$p :\I \fun (x:\Gamma)\times (A\,x\times A\,x)$, $\varphi :\Cof$,
$f :[\varphi]\fun \Pi_\I((\Path A)\comp p)$, $q :\Path A\,(p\,e)$ with
$(\varphi,f)\conc e \exto q$ and $i :\I$, suppose
$p=\Pair{p'}{\Pair{q_0}{q_1}}$ where $p':\I\fun\Gamma$ and
$q_0, q_1:\Pi_\I(A\comp p)$, and define
\begin{equation}
    \isFib_{\Path}\alpha\,e\,p\,\varphi\,f\,q\,i \defeq \alpha\,e\,p'\,(\varphi
    \disj{i=\src}\disj{i=\tgt})\,(f'\join f_0\join f_1)\,(q\,i)
    \label{eq:54}
  \end{equation} 
where
\begin{align*}
  &f' :[\varphi] \fun\Pi_\I(A\comp p')\\
  &f'\,u\,j \defeq f\,u\,j\,i\\
  &f_0 : \{g:[i=\src] \fun \Pi_\I(A\comp p') \mid
    (\varphi,f')\compat ({i=\src},g)\}\\
  &f_0\,\_ \defeq q_0\\
  &f_1 : \{g:[i=\tgt]\fun \Pi_\I(A\comp p') \mid
    (\varphi \disj{i=\src},f' \join f_0)\compat
    ({i=\tgt},g)\}\\ 
  &f_1\,\_ \defeq q_1
\end{align*}
Thus for each $i:\I$ we have
$\isFib_{\Path}\alpha\,e\,p\,\varphi\,f\,q\,i :A(p'\inv{e})$, so that
$\isFib_{\Path}\alpha\,e\,p\,\varphi\,f\,q :\I\fun A(p'\inv{e})$. Since
$\alpha\,e\,p':\Comp e\,(A\comp p')$, we have
\begin{multline*}
  \forall(u:[\varphi \disj{i=\src}\disj{i=\tgt}]).\;
  (f'\join f_0\join f_1)\,u\,\inv{e} = \\
  \alpha\,e\,p'\,(\varphi \disj{i=\src}\disj{i=\tgt})\, (f'\join
  f_0\join f_1)\,(q\,i) = \isFib_{\Path}\alpha\,e\,p\,\varphi\,f\,q\,i
\end{multline*}
Hence $\isFib_{\Path}\alpha\,e\,p\,\varphi\,f\,q\,\src = q_0\,\inv{e}$ and
$\isFib_{\Path}\alpha\,e\,p\,\varphi\,f\,q\,\tgt = q_1\,\inv{e}$, so that
\[
\isFib_{\Path}\alpha\,e\,p\,\varphi\,f\,q: \Path A\,(p'\,\inv{e},
(q_0\,\inv{e},q_1\,\inv{e})) = \Path A\,(p\,\inv{e})
\]
and furthermore,
$(\varphi,f)\conc\inv{e} \exto
\isFib_{\Path}\alpha\,e\,p\,\varphi\,f\,q$.
Thus $\isFib_{\Path}\alpha : \isFib(\Path A)$. Finally, property
\eqref{eq:28} is immediate from \eqref{eq:54} and the definition of
$\_[\_]$~\eqref{eq:21}.  
\qed

These path types in the CwF of CCHM fibrations
(Definition~\ref{def:cchm-fib-cwf}) satisfy the Coquand formulation of
identity types with propositional computation
properties~\cite[Figure~2]{CoquandT:modttc}. Thus in addition to the
contractibility of singleton types~\eqref{eq:6}, we get
\emph{substitution functions} for transporting elements of a fibration
along a path
\begin{align}
  &\kw{subst}: \{\Gamma:\Univ\}\{A:\Gamma\fun\Univ\}\{\alpha:\isFib A\}
  \{x_0\;x_1:\Gamma\} \fun
  (x_0\path x_1) \fun A\,x_0\fun A\,x_1\label{eq:90}\\
  &\kw{subst}\,p\,a \defeq \alpha\,\src\,p\,\False\,\elim_\emptyset\,a
  \notag 
\end{align}
using the cofibrant partial elements $(\False, \elim_\emptyset)$
mentioned after Definition~\ref{def:cpe}. By
Lemma~\ref{lem:fill} we have that these substitution functions satisfy
a propositional computation rule for constant paths~\eqref{eq:5}:
\begin{align}
  &\kw{H} :
    \{\Gamma:\Univ\}\{A:\Gamma\fun\Univ\}\{\alpha:\isFib A\}
    \{x:\Gamma\}(a : A\,x) \fun (a \path {\kw{subst}(\idp x)\,a})
    \label{eq:91}\\
    &\kw{H}\,a \defeq \notag \fil\src\,\alpha\,(\idp
    x)\,\False\,\elim_\emptyset\,a
\end{align}

\begin{rem}[\textbf{Function extensionality}]
  \label{rem:funext}
  As one might expect from~\cite[Lemma~6.3.2]{HoTT}, the path types of
  Theorem~\ref{thm:fib-path} satisfy function extensionality. Given
  $A:\Univ$, $B:A\fun \Univ$, $f,g:(x:A)\fun B\,x$ and $p:(x:A)\fun
  (f\,x\path g\,x)$, we get a path $\kw{funext}\,p:f\path g$ in
  $(x:A)\fun B\,x$ given by
  \[
  \kw{funext}\,p\,i \defeq \lambda(x:A)\fun p\,x\,i 
  \]
  for all $i:\I$.
\end{rem}

To get Martin-L\"of identity types with standard definitional, rather
than propositional computation properties from these path types, we
use a version of Swan's construction~\cite{SwanA:algwfs} like the one
in Section~9.1 of~\cite{CoquandT:cubttc}, but only using the
connection algebra structure on $\I$, rather than a de~Morgan
algebra structure. This is the only place that axiom $\pos{7}$ is
used; we need the fact that the universe given by $\Cof$ and
$[\_]:\Cof\fun\Univ$ is closed under dependent products:
\begin{lem}
  \label{lem:cofpdp}
  The following element of type $\Prop$ is provable:
  $\forall(\varphi:\Prop)(f:[\varphi]\fun\Prop). \cof \varphi \imp
  (\forall (u:[\varphi]).\; \cof(f\,u)) \imp
  \cof(\exists(u:[\varphi]).\; f\,u)$.
\end{lem}
\proof
  Note that if $u:[\varphi]$ then
  $(\exists(v:[\varphi]).\;f\,v) = f\,u$ and hence
  $\cof(\exists(v:[\varphi]).\;f\,v) = \cof(f\,u)$.  So
  $\forall (u:[\varphi]).\; \cof(f\,u)$ equals
  $\varphi \imp \cof(\exists(v:[\varphi]).\;f\,v)$. Therefore from
  $\cof\varphi$ and $\forall (u:[\varphi]).\; \cof(f\,u)$ by axiom
  $\pos{7}$ we get $\cof(\varphi \conj \exists(v:[\varphi]).\;f\,v)$
  and hence $\cof(\exists(v:[\varphi]).\;f\,v)$, since
  $(\exists(v:[\varphi]).\;f\,v) \imp \varphi$.
\qed

\begin{thm}[\textbf{Fibrant identity types}]
  \label{thm:fib-id}
  Define identity types by:
  \begin{align}
    &{\Id} : \{\Gamma:\Univ\}(A:\Gamma\fun\Univ) \fun (x:\Gamma)\times
    (A\,x\times A\,x) \fun \Univ\label{eq:2}\\
    &\Id A\,(x,(a_0,a_1)) \defeq 
    (p:\Path A\,(x,(a_0,a_1))) \times \{ \varphi:\Cof\mid
    \varphi \imp \forall(i:\I).\;p\,i=a_0\}\notag
  \end{align}
  Then there is a function
  $\isFib_{\Id} : \{\Gamma:\Univ\}\{A:\Gamma\fun\Univ\} \fun \isFib A
  \fun \isFib (\Id A)$
  and the fibrations $(\Id A , \isFib_{\Id}A)$ can be given the
  structure of Martin-L\"of identity types in the CwF of CCHM
  fibrations~\cite[Defintion~3.19]{HofmannM:synsdt}.
\end{thm}
\proof
Given $\Gamma :\Univ$, $A :\Gamma\fun\Univ$ and $\alpha :\isFib A$,
using Theorems~\ref{thm:fib-sigma} and~\ref{thm:fib-path} we 
define
$\isFib_{\Id}\alpha \defeq
\isFib_{\Sigma}(\isFib_{\Path}\alpha)\,\beta$, where $\beta
:\isFib\Phi$ with
\begin{align*}
  &\Phi : (y : (x:\Gamma) \times (A\,x \times A\,x)) \times \Path A\,y \fun
    \Univ\\ 
  &\Phi((x,(a_0,a_1)),p) \defeq \{\varphi:\Cof \mid
    \varphi \imp \forall(i:\I).\;p\,i = a_0\}
\end{align*}
and the fibration structure $\beta$ mapping $e : \Bool$,
$p:\I\fun (y : (x:\Gamma) \times (A\,x \times A\,x)) \times \Path
A\,y$,
$\varphi:\Cof$, $f : [\varphi] \fun \Pi_\I(\Phi\comp p)$ and
$\varphi':\Phi(p\,e)$ with $(\varphi,f)\conc e \exto \varphi'$ to
the element
\[
\beta\,e\,p\,\varphi\,f\,\varphi' \defeq \exists(u:[\varphi]).\;
f\,u\,\inv{e}
\]
(using Lemma~\ref{lem:cofpdp} to see that this is well defined).  We
get the usual introduction, elimination and computation rules for
these identity types as follows.  Since $\True:\Cof$ holds by axiom
$\pos{5}$, identity introduction
\begin{equation}
  \label{eq:29}
  {\refl} : \{\Gamma:\Univ\}\{A:\Gamma\fun\Univ\}\{x:\Gamma\}(a :
  A\,x) \fun \Id A\,(x,(a,a))
\end{equation}
can be defined by $\refl a \defeq (\lambda a\;i\fun a , \True)$.
Identity elimination
\begin{equation}
  \label{eq:34} 
  {\kw{J}}: 
  \begin{array}[t]{@{}l}
    \{\Gamma:\Univ\}(A:\Gamma\fun\Univ)(x:\Gamma)(a_0:A\,x)
    (B: (a:A\,x) \times \Id A\,(x,(a_0,a))\fun \Univ)\\
    (\beta:\isFib B)(a_1:A\,x)(e :\Id A\,(x,(a_0,a_1))) \fun
    B(a_0,\refl a_0) \fun B(a_1,e)
  \end{array} 
\end{equation}
is given by
\[
\kw{J}\,A\,x\,a_0\,B\,\beta\,a_1(p,\varphi)\,b \defeq \beta\,\src\,
\Pair{p}{q}\,\varphi\,f\,b
\]
where $q:(i:\I)\fun \Id A\,x\,(a_0, p\,i)$ is
$q\,i\,j \defeq (p(i \sqcap j), \varphi\disj{i=\src})$ and
$f:[\varphi]\fun\Pi_\I(B\comp\Pair{p}{q})$ is $f\,u\,i \defeq b$.  (In
the above element, since $(p,\varphi): \Id A\,(x,(a_0,a_1))$ we have
$p\,\src=a_0$, $p\,\tgt=a_1$ and
$\varphi\imp\forall(i:\I).\;p\,i=a_0$; hence
$\varphi \imp \forall(i:\I).\;{q\,i = \refl a_0}$, so that $f$ is
well-defined.) Note that by axioms $\pos{3}$ and $\pos{4}$ we have
$q\,\src = \refl a_0$ and $q\,\tgt = (p,\varphi)$, so that
$\kw{J}\,A\,x\,a_0\,B\,\beta\,a_1(p,\varphi)\,b = \beta\,\src\,
\Pair{p}{q}\,\varphi\,f\,b : B(p\,\tgt,q\,\tgt) =
B(a_1,(p,\varphi))$,
as required. Furthermore, since
$(\varphi,f)\conc\tgt \exto \beta\,\src\,
\Pair{p}{q}\,\varphi\,f\,b$, we have
\[
\forall(u:[\varphi]).\; b = f\,u\,\tgt =
\kw{J}\,A\,x\,a_0\,B\,\beta\,a_1(p,\varphi)\,b
\]
So when $(p,\varphi)= \refl a_0 = (\idp a_0,\True)$ and hence
$a_1=a_0$, we have 
\begin{equation}
  \label{eq:30}
  \kw{J}\,A\,x\,a_0\,B\,\beta\,a_o\,(\refl a_0)\,b = b
\end{equation}
In other words the computation property for identity elimination
holds as a judgemental equality and not just a propositional one.
Finally, to correctly support the interpretation of intensional
identity types, one needs stability of $(\Id A , \isFib_{\Id}A)$,
$\refl$ and $\kw{J}\,A$ under re-indexing; but this follows from the
stability properties of $\isFib_{\Sigma}$ and $\isFib_{\Path}$.  \qed

\section{Glueing}
\label{sec:glueing}

In this section we give an internal presentation of the \emph{glueing}
construction given by Cohen~\emph{et
  al}~\cite{CoquandT:cubttc}. Glueing is similar to a composition
structure (Definition~\ref{def:cchm-fib}) for type-families, except
that instead of partial paths of types it involves partial
equivalences between types. Glueing is crucial for the constructions
relating to univalence~\cite{HoTT} given in Section \ref{univalence}.

We begin by defining the glueing construction for cofibrant-partial
types, that is, for functions $A:[\varphi] \fun \Univ$ where
$\varphi:\Cof$:

\begin{defi}[\textbf{Glueing}]
\label{defi:glu}
  Given
  $\varphi:\Cof$,
  $A:[\varphi] \fun \Univ$,
  $B:\Univ$ and
  $f:(u:[\varphi])\fun A\,u \fun B$, the type $\Glue\varphi\,A\,B\,f :
  \Univ$ is defined to be
  \begin{align}
    \Glue\varphi\,A\,B\,f 
    &\defeq (a:(u:[\varphi])\fun A\,u) \times
      \{b :B \mid \forall(u:[\varphi]).\; f\,u\,(a\,u) =
      b\}
  \end{align}
  Elements of this type consist of pairs $(a,b)$ where $a$ is a partial
  element of the partial type $A$ and $b$ is an element of type $B$,
  such that $f$ applied to $a$ gives a partial element of $B$ that
  extends to $b$. When $\varphi = \True$ then $A$ and $f$ are both total
  and so $\Glue\varphi\,A\,B\,f$ essentially consists of pairs
  $(a,f\,a)$ for every $a : A$ and hence is clearly isomorphic to
  $A$. When $\varphi = \False$ then $A$ and $f$ are both uniquely
  determined and $\Glue\varphi\,A\,B\,f$ will consist of pairs
  $(\elim_\emptyset,b)$ for $b : B$ and hence is clearly isomorphic to
  $B$.
\end{defi}

We now extend this glueing operation from cofibrant-partial types to
\emph{cofibrant-partial type-families}:

\begin{defi}[\textbf{Cofibrant-partial families}]
  \label{defi:cofpf}
  Given a object $\Gamma : \Univ$ and a cofibrant property
  $\Phi : \Gamma \fun \Cof$ define the \emph{restriction of $\Gamma$
    by $\Phi$} to be
  \begin{equation}
    \label{eq:73} 
    \Gamma\resby \Phi \defeq (x:\Gamma)\times[\Phi\,x]
  \end{equation}
  Thus $\Gamma\resby \Phi : \Univ$ and there is a monomorphism
  $\iota:\Gamma\resby\Phi \mono \Gamma$ given by first projection.
  (Note that $\Gamma\resby \Phi$ is isomorphic to the comprehension
  subtype $\{x:\Gamma\mid\Phi\,x\}$, but we use the above
  representation to make proofs of $\Phi\, x$ more explicit in various
  constructions.)  Then given an object $\Gamma : \Univ$ and a
  cofibrant property $\Phi : \Gamma \fun \Cof$, a
  \emph{cofibrant-partial type-family} over $\Gamma$ is a family $A$
  of types over the restriction $\Gamma\resby \Phi$, that is
  $A : (\Gamma\resby\Phi) \fun \Univ$.
\end{defi} 

\begin{defi}[\textbf{Glueing for families}]
  \label{defi:gluf}
  We lift the glueing operation from types to type-families as
  follows.  Given $\Gamma:\Univ$, $\Phi:\Gamma\fun\Cof$,
  $A:\Gamma\resby\Phi \fun \Univ$, $B:\Gamma\fun\Univ$ and
  $f:(x:\Gamma)(v:[\Phi\,x])\fun A(x,v) \fun B\,x$, define the family
  $\Glue\Phi\,A\,B\,f : \Gamma \fun \Univ$ by
  \begin{align}
    \Glue\Phi\,A\,B\,f\,x 
    &\defeq\Glue\,(\Phi\,x)\,(A(x,\_))\,(B\,x)\,(f(x,\_)) 
  \end{align}
\end{defi} 

The glueing construction works for any map
$f:(x:\Gamma)(v:[\Phi\,x])\fun A(x,v) \fun B\,x$. However, we want to
see that this construction lifts to the CwF of CCHM fibrations. This
means that $\Glue\Phi\,A\,B\,f$ should have a fibration structure
whenever $A$ and $B$ do and this puts some requirements on $f$. We
begin by introducing the notion of an \emph{extension structure}:

\begin{defi}[\textbf{Extension structures}]
  The type of extension structures, ${\Ext} : \Univ \fun \Univ$, is
  given by
  \[
  \Ext A \defeq (\tilde{a} : \Cpf A) \fun \{a:A\mid \tilde{a}\exto a\}
  \]
  Having an extension structure for a type $A:\Univ$ allows us to extend
  any partial element of $A$ to a total element. We say that a family
  $A:\Gamma\fun\Univ$ has an extension structure if each of its fibres
  do, and we abusively write
  \[ 
  \Ext A \defeq (x : \Gamma) \fun \Ext (A\, x)
  \]
\end{defi}
An extension structure for $A:\Gamma\fun\Univ$ is similar to having a
composition structure for $A$ in the sense that both allow us to
extend partial elements; and in fact every family with an extension
structure is a fibration.  However, an extension structure does not
require a total element from which we extend/compose and so is in fact
a stronger notion than a composition structure.  First note that
having an extension structure for $A:\Univ$ implies that $A$ is
inhabited, because we can always extend the empty partial
element. Further, given any element $a : A$ we can use the extension
structure to show that it is path equal to the extension of the empty
partial element. Together these facts tell us that $A$ is
\emph{contractible}:

\begin{defi}[\textbf{Contractibility~\cite[Section~3.11]{HoTT}}]
  A type $A$ is said to be \emph{contractible} if it has a centre of
  contraction $a_0 : A$ and every element $a : A$ is propositionally
  equal to $a_0$, that is, there exists a path $a_0 \path a$.
  Therefore a type is contractible if $\Contr A$ is inhabited, where
  ${\Contr} : \Univ \fun \Univ$ is defined by
  \[ 
  \Contr A \defeq (a_0 :A)\times((a:A)\fun {a_0\path a}) 
  \]
  As with extension structures, we say that a family
  $A:\Gamma\fun\Univ$ is contractible if each of its fibres is and
  write
  \[ 
  \Contr A \defeq (x:\Gamma)\fun \Contr(A\,x) 
  \]
\end{defi}
As mentioned above, having an extension structure for a family
$A:\Gamma\fun\Univ$ implies that $A$ is both fibrant and
contractible. In fact the converse is true as well
(\emph{cf.}~\cite[Lemma~5]{CoquandT:cubttc}):
\begin{lem}
\label{lem:ext}
  There are functions
  \begin{gather}
    \kw{fromExt}:\{\Gamma:\Univ\}\{A:\Gamma\fun
    \Univ\}\fun \Ext A \fun \isFib A \times \Contr A\\
    \kw{toExt}: \{\Gamma:\Univ\}\{A:\Gamma\fun
    \Univ\}\fun \isFib A \fun \Contr A \fun
    \Ext A \label{eq:13}
  \end{gather}
\end{lem}
\proof
  Given $\Gamma:\Univ, A:\Gamma\fun\Univ$ and $\varepsilon : \Ext A$
  we define $\alpha : \isFib A$
  \[
  \alpha\; e\; p\;\varphi\;f\;a_0 \defeq
  \varepsilon\;(p\;\inv{e})\;((\varphi, f)\conc \inv{e})
  \]
  For every $x:\Gamma$ we use the totally undefined cofibrant-partial
  element $(\False,\elim_\emptyset):\Cpf A$ to define $a_0 : A$
  \begin{gather}
    a_0 \defeq
    \varepsilon\,x\,(\False,\elim_\emptyset)\notag  
  \end{gather}
  For each $a:A\,x$ and $i:\I$, we have $\lambda\_\fun a : [i=\tgt]\fun
  A\,x$; so we get a path $p_a : \I \fun A\,x$
  \[
  p_a\, i \defeq \varepsilon\,x\,({i=\tgt},\lambda\_\fun a)
  \]
  By the definition of $\Ext A$ we have
  $p_a\,\tgt = \varepsilon\,x\,(\True,\lambda\_\fun a) = a$,
  and by $\pos{2}$ we have
  $p_a\,\src = \varepsilon\,x\,(\False,\elim_\emptyset) = a_0$.
  Therefore $p_a:a_0 \path a$. Hence $A\,x$ is contractible. 
  Together this shows that, given $\varepsilon : \Ext A$, we can
  define elements of type $\isFib A$ and $\Contr A$.  Therefore there is
  a function
  $\kw{fromExt}:\{\Gamma:\Univ\}\{A:\Gamma\fun \Univ\}\fun \Ext A
  \fun \isFib A \times \Contr A$.

  Conversely, given $\Gamma :\Univ$, $A :\Gamma\fun \Univ$,
  $\alpha :\isFib A$, $\langle a_0, p\rangle : \Contr A$, note that for any
  $x:\Gamma$, $\varphi :\Cof$ and $ f: [\varphi]\fun A\,x$ we have
  $(p\,x)\comp f : [\varphi]\fun(\I\fun A\,x)$ such that
  $\forall(u:[\varphi]).\; ((p\,x)\comp f)\,u : a_0\,x \path f\,u$;
  therefore $(\varphi,(p\,x)\comp f)\conc\src \exto (a_0\,x)$ and so
  defining
  \[
   \varepsilon\,x\,(\varphi,f) \defeq
   \alpha\,\src\,(\lambda\_\fun x)\,\varphi\,((p\,x)\comp f)\,(a_0 \,x)
  \]
  we get $\varepsilon\,x\,(\varphi,f) : A\,x$.
  Furthermore, since
  $(\varphi,(p\,x)\comp f)\conc\tgt = (\varphi,f)$ by the 
  type of $\alpha$ we get $(\varphi,f) \exto \varepsilon\,x\,(\varphi,f)$.
  Thus $\varepsilon\,x\,(\varphi,f) : \Ext(A\,x)$, as required. 
\qed

We now come to the main result of this section: showing that
fibrations are closed under glueing. Proving this requires that the
function $f$ is an \emph{equivalence}:

\begin{defi}[\textbf{Equivalences~\cite[Section~4.4]{HoTT}}]
  Given types $A,B:\Univ$, a function $f : A \fun B$ is an
  \emph{equivalence} if the type $\Equiv\, f$ is inhabited, where
  \[
  \Equiv f \defeq (b:B) \fun \Contr((a:A)\times{f\,a \path b})
  \]
  Again, this lifts to families in the obvious way: given
  $\Gamma:\Univ$, $A,B:\Gamma\fun\Univ$ and
  $f : (x:\Gamma)\fun A\,x\fun B\,x$, define
  \[
  \Equiv f \defeq (x : \Gamma) \fun \Equiv\,(f\,x)
  \]
\end{defi}

\begin{thm}[\textbf{Composition for glueing}]
  \label{thm:fib-glue}
  Let $\Phi$, $A$, $B$ and $f$ be as in
  Definition~\ref{defi:gluf}. Then $\Glue\Phi\,A\,B\,f$ has a
  fibration structure if $A$ and $B$ both have one and $f$ has the
  structure of an equivalence. In other words there is a function 
  \begin{equation}
    {\isFib_{\Glue}} :
    \begin{array}[t]{@{}l}
      \{\Gamma:\Univ\}
      \{\Phi:\Gamma\fun\Cof\}
      \{A:\Gamma\resby\Phi \fun \Univ\}
      \{B:\Gamma\fun\Univ\}\\
      (f:(x:\Gamma)(u:[\Phi\,x])\fun A(x,u) \fun B\,x)\fun{}\\
      ((x:\Gamma)(v:[\Phi\,x]) \fun \Equiv(f\,x\,v)) \fun{}\\
      \isFib A \fun \isFib B \fun \isFib(\Glue\Phi\,A\,B\,f)
    \end{array}\label{eq:88}
  \end{equation}
\end{thm}
\proof
  Given $\Gamma:\Univ$, $\Phi:\Gamma\fun\Cof$,
  $A:\Gamma\resby\Phi \fun \Univ$, $B:\Gamma\fun\Univ$,
  $f:(x:\Gamma)(u:[\Phi\,x])\fun A(x,u) \fun B\,x$,
  $eq:(x:\Gamma)(u:[\Phi\,x]) \fun \Equiv(f\,x\,v)$,
  $\alpha:\isFib A$, $\beta:\isFib B$, we wish to define an element of
  type $\isFib(\Glue\Phi\,A\,B\,f)$. Therefore, taking
  \begin{gather*}
  e :\Bool,\;
  p : \I \fun \Gamma,\;
  \psi : \Cof,\;
  q:[\psi]\fun\Pi_\I(\Glue\Phi\,A\,B\,f)\\
  (a_0,b_0):\{ (a_0,b_0) : (\Glue\Phi\,A\,B\,f)(p\,e) \mid (\psi,q)
  \conc e \exto (a_0,b_0)\} 
  \end{gather*}
  our goal is to define $(a_1,b_1) : (\Glue\Phi\,A\,B\,f)\,x$ such
  that $(\psi,\tilde{a_1}) \exto a_1$ and
  $(\psi,\tilde{b_1}) \exto b_1$, where
  $x : \Gamma$, $\tilde{a_1}:[\psi]\fun ((u:[\Phi\,x])\fun A(x,u))$
  and $\tilde{b_1}:[\psi]\fun B\;x$ are defined by
  \begin{align*}
  x &\defeq p\, \inv{e}\\
  \tilde{a_1}\,v &\defeq \fst(q\; v\; \inv{e})\\
  \tilde{b_1}\,v &\defeq \snd(q\; v\; \inv{e})
  \end{align*}
  and satisfy
  $\forall(v:[\psi])(u:[\Phi\,x]).\;f\,x\,u\,(\tilde{a_1}\,v\,u) =
  \tilde{b}\, v$ by the definition of $\Glue\Phi\,A\,B\,f$.
  
  We start by composing over $p$ in
  $B$ to get $b'_1 : B\;x$ which can be thought of as a first
  approximation to $b_1$: 
  \[
  b'_1 \defeq \beta\; e\; p\;\psi\;(\lambda(v:[\psi])(i:\I) \fun
  \snd(q\,v\,i))\; b_0
  \]
  Recall that $a_1$ will have type $(u:[\Phi\; x]) \fun A(x,u)$ and so
  we assume $u : [\Phi\,x]$ in order to define an element of type
  $A(x,u)$. Note that we cannot simply compose over $p$ in $A$ because
  we do not know that $\Phi(p\, i)$ holds for all $i : \I$. Instead we
  will use the equivalence structure to define $a_1$.
  
  Let $C \defeq (a : A(x,u)) \times f\,x\,u\,a\path b'_1$ be the fibre
  of $f\,x\,u$ at $b'_1$. Using Theorems \ref{thm:fib-sigma} and
  \ref{thm:fib-path} and the fact that both $A$ and $B$ are fibrations
  (as witnessed by $\alpha$ and $\beta$ respectively) we can deduce
  that $C$ is a fibrant object. Combined with the fact that
  $eq\;x\;u\;b'_1 : \Contr C$ we can use Lemma \ref{lem:ext} to define
  $\varepsilon : \Ext C$. We can then define
  \[ 
  (\psi,(\lambda(v:[\psi]) \fun\tilde{a_1}\,v\,u,\; \refl \comp\;
  \tilde{b_1})) : \Cpf C
  \]
  This is well-defined because, as mentioned above, we have
  $\forall(v:[\psi]).\;f\;x\;u\;(\tilde{a_1}\,v\,u) = \tilde{b_1}\;v$
  and, by the type of composition structures, we have
  $(\psi,\tilde{b_1}) \exto b'_1$ and so given $v:[\psi]$ we have
  $\refl (\tilde{b_1}\,v) : f\;x\;u\;(\tilde{a_1}\,v\,u) \path
  b'_1$. Now we can define
  \[
  \varepsilon\,(\psi,(\lambda(v:[\psi]) \fun\tilde{a_1}\,v\,u,\; \refl
  \comp\; \tilde{b_1})) : C 
  \]
  Discharging our assumption $u:[\Phi\,x]$ and taking first and second
  projections of the pair defined above we get:
  \[
  a_1 : (u:[\Phi\,x]) \fun A(x,u) \qquad p_b : (u:[\Phi\,x]) \fun
  f\,x\,u\,(a_1\,u) \path b'_1 
  \]  
  We now have $a_1$ and $b'_1$ such that $\tilde{a_1}\exto a_1$ and
  $\tilde{b_1}\exto b'_1$. However, we cannot simply take $b_1$ to be
  $b'_1$ because we do not know that
  $\forall(u:\Phi\;x).\,f\;x\;u\;(a_1\;u) = b'_1$ and therefore cannot
  conclude that $(a_1,b'_1):(\Glue\Phi\,A\,B\,f)\,x$. In order to
  solve this problem we perform one final composition in $B\;x$ in
  order to ``correct" $b'_1$ to achieve this property. Consider the
  following join
  \[ 
  p_b \join (\refl \comp\; \tilde{b_1}) : [\Phi\,x \disj \psi] \fun
  \Pi_\I(B\,x) 
  \]
  This is well defined because $p_b$ is defined by extending
  $\refl \comp\; \tilde{b_1}$ and so they must be equal where they are
  both defined. We use this to perform one final composition in
  $B\,x$:
  \[ 
  b_1 \defeq \beta\;\tgt\;(\lambda(\_:\I)\fun x)\;(\Phi\,x \disj
  \psi)\;(p_b \join (\refl \comp\; \tilde{b_1}))\;b'_1
  \]
  We now have $(a_1,b_1) : (\Glue\Phi\,A\,B\,f)\,x$ such that
  $(\psi,\tilde{a_1}) \exto a_1$ and $(\psi,\tilde{b_1}) \exto b_1$,
  as required.  
\qed

We now have a way to interpret the glueing operation from
\cite{CoquandT:cubttc} that meets some of the necessary requirements;
see~\cite[Figure 4]{CoquandT:cubttc}. However, the current
construction fails the requirement that $\Glue\Phi\,A\,B\,f$ should be
equal to $A$ when reindexing along the inclusion
$\iota : \Gamma\resby\Phi \mono \Gamma$. In fact, this equality should
hold in the CwF of CCHM fibrations. This means that not only should
$A = (\Glue\Phi\,A\,B\,f)\comp \iota : \Gamma\resby\Phi \to \Univ$,
but also that reindexing the fibration structure derived in Theorem
\ref{thm:fib-glue} should result in the same fibration structure with
which we started. To be precise, what we require is:
\begin{equation}
  \label{eq:35}
  (A,\alpha) = (\Glue\Phi\,A\,B\,f, \isFib_{\Glue}\,f\,eq\,\alpha\,\beta) [\iota]
\end{equation}
What we have at present is that the families $A$ and
$(\Glue\Phi\,A\,B\,f) \comp \iota$ are isomorphic in the following sense:

\begin{defi}[\textbf{Isomorphisms}]
  Given objects $A,B:\Univ$, an isomorphism between $A$ and $B$ is a
  function $f:A \fun B$ that has a 2-sided inverse. Let $A\cong B$ be
  the type of isomorphisms between $A$ and $B$, defined by
  \[
  A \cong B \defeq \{ f : A \to B \mid (\exists g : B \fun A)\, (g
  \comp f = id) \conj (f \comp g = id) \}
  \]
  We say that $A$ and $B$ are isomorphic if there exists an
  isomorphism $f:A\cong B$. This notion lifts to families in the
  obvious, point-wise way.  Isomorphisms have inverses up to the
  extensional equality of the internal type theory, in contrast to
  equivalences which only have inverses up to path equality.
\end{defi}

We get to \eqref{eq:35} in two steps. First we use Axiom~$\pos{9}$ in
order to \emph{strictify} the glueing construction to get a new,
strict form of glueing, $\SGlue$, such that
$\Glue\Phi\,A\,B\,f \cong \SGlue\Phi\,A\,B\,f$ but where
$A = (\SGlue\Phi\,A\,B\,f)\comp \iota$. We then use Axiom~$\pos{8}$ to
adapt the fibration structure on $\SGlue$ so that under reindexing
along $\iota$ it is equal the the fibration structure on $A$. The
order of these steps does not seem to be important; we could equally
have first adapted the fibration structure on $\Glue$ and then
strictified $\Glue$ with this new fibration structure.  

\begin{rem}
  Apart from the use of the internal language of a topos, our approach
  to getting a glueing operation with good properties diverges from
  that taken by Cohen~\emph{et al} \cite{CoquandT:cubttc}, where
  glueing is defined directly with all the required
  properties. However it is possible to see where our final two steps
  occur in the original work. The strictification can be seen in the
  use of the case split on $\varphi\rho=1_{\mathbb{F}}$ in
  \cite[Definition 15]{CoquandT:cubttc}; see
  Section~\ref{sec:sat-strictness} for more details. Rather than
  defining an initial composition structure for glueing and then
  modifying it to get the required reindexing property, Cohen~\emph{et
    al} define the composition structure directly. Removing all uses
  of the $\forall$ operator from \cite[Section 6.2]{CoquandT:cubttc}
  would yield our initial composition structure, and we then use the
  closure of $\Cof$ under $\forall(i:\I)$ (axiom $\pos{8}$) in a
  separate step to modify this composition. We prefer this approach
  because it simplifies the core composition structure for glueing and
  makes more explicit what role $\pos{8}$ plays in the construction of
  a model of cubical type theory.
\end{rem}

We now recall axiom $\pos{9}$ from Figure~\ref{fig:axi}:
\[
\strictify :
\begin{array}[t]{@{}l}
  \{ \varphi : \Cof \}
  (A: [\varphi] \fun \Univ)
  (B: \Univ)
  (s : (u : [\varphi]) \fun (A\, u \cong B))
  {}\fun{}\\
  (B' : \Univ) \times \{s' : B' \cong B \mid \forall(u :
  [\varphi]).\; A\, u = B' \conj s\,u = s'\}
\end{array}
\]  
This states that any partial type $A$, which is isomorphic to a
total type $B$ everywhere that it is defined, can be extended to a
total type $B'$ that is isomorphic to $B$. We investigate why the
cubical presheaf topos~\cite{CoquandT:cubttc} satisfies this axiom
in section~\ref{sec:sat-strictness}.  Given $\pos{8}$, it is
straightforward to define a strict form of glueing.

\begin{defi}[\textbf{Strict glueing}]
  \label{defi:strg}
  Given $\Gamma:\Univ$, $\Phi:\Gamma\fun\Cof$, $A:\Gamma\resby\Phi
  \fun \Univ$, $B:\Gamma\fun\Univ$ and $f:(x:\Gamma)(u:[\Phi\,x])\fun
  A(x,u) \fun B\,x$, define $\SGlue \Phi\,A\,B\,f : \Gamma\fun \Univ$
  by 
  \begin{gather}
    \SGlue\Phi\,A\,B\,f\,x \defeq \hspace{320pt}\notag\\
    \qquad  \fst(\strictify\, (\lambda u : [\Phi\, x]
      \fun A(x,u))\, (\Glue\,\Phi\,A\,B\,f\, x)\, (\lambda u : [\Phi\,
      x] \fun \glue\,(x,u))
  \end{gather}	
  where $\glue\, (x,u) : A(x,u) \cong \Glue\,\Phi\,A\,B\,f\, x$ is the
  isomorphism alluded to in Definition~\ref{defi:glu} given by
  \[ \glue\; (x,u)\; a \defeq (\lambda \_ \fun a, f\,x\,u\,a) \]
  Note that $\SGlue$ has the desired strictness property: given any
  $(x,u) : \Gamma\resby\Phi$, by $\pos{9}$ we have
  $A(x,u) = \fst(\strictify\, (\lambda u : [\Phi\, x] \fun A(x,u))\,
  (\Glue\,\Phi\,A\,B\,f\, x)\, (\lambda u : [\Phi\, x] \fun \glue\,
  (x,u)))$ and hence
  \begin{equation}
    \forall(x:\Gamma)(u:[\Phi\,x]).\; \SGlue\Phi\,A\,B\,f\,x = A(x,u)
  \end{equation}
\end{defi}

\begin{thm}[\textbf{Composition structure for strict glueing}]
  \label{thm:fib-sglue}
  Given $\Gamma$, $\Phi$, $A$, $B$, and $f$ as in
  Definition~\ref{defi:strg}, $\SGlue\Phi\,A\,B\,f :\Gamma\fun \Univ$
  has a fibration structure if $A$ and $B$ have one and $f$ has the
  structure of an equivalence.
\end{thm}
\proof
  It is easy to show that (fibrewise) isomorphisms preserve fibration
  structures. Hence we obtain a fibration structure on $\SGlue$ by
  transporting the structure obtained from $\isFib_{\Glue}$
  (Theorem~\ref{thm:fib-glue}) along the isomorphism from
  $\strictify$.
\qed

The final step in this section in this section is to use
axiom~$\pos{8}$ to adapt the composition structure for $\SGlue$ so
that we recover the original composition structure on $A$.

\begin{thm}[\textbf{Adapting composition structures}]
  \label{thm:adapt}
  Given $\Gamma:\Univ$ and $\Phi:\Gamma\fun\Cof$, let
  $\iota:\Gamma\resby\Phi \mono \Gamma$ be as in
  Definition~\ref{defi:cofpf}. For any $G:\Gamma\fun \Univ$,
  $\alpha : \isFib(G \comp \iota)$ and $\gamma : \isFib{G}$, there
  exists a composition structure $\gamma' : \isFib{G}$ such that
  $\alpha = \gamma'[\iota]$.
\end{thm}
\proof Given $\Gamma,\Phi,A,\alpha,\gamma$ as above, using axiom
$\pos{8}$ (and $\pos{6}$) we can define $\gamma'$ by
  \begin{equation}
    \gamma'\,e\,p\,\psi\,f\,g \defeq
    \gamma\,e\,p\,(\psi\disj(\forall(i:\I).\,\Phi\,(p\,i)))\,(f
    \cup f')\,g
  \end{equation}
  where $f' : [\forall(i:\I).\,\Phi\,(p\,i)] \fun \Pi_\I G$ is given
  by
  $f'\,u \defeq \fil \,e\,\alpha\,(\lambda i \fun (p,
  u\,i))\,\psi\,f\,g$. The fact that $f$ and $f'$ are compatible
  follows from the fact that we use $\psi$ and $f$ in the definition
  of $f'$ and so by the properties of filling $f'$ must agree with
  $f$ wherever they are mutually defined. \qed

\begin{cor}\label{cor:glue}
  Given $\Gamma:\Univ$, $\Phi:\Gamma\fun\Cof$,
  $(A,\alpha):\Fib(\Gamma\resby\Phi)$, $(B,\beta):\Fib\Gamma$ and
  $f:(x:\Gamma)(v:[\Phi\,x])\fun A(x,v) \fun B\,x$, there exists
  $(G,\gamma):\Fib\Gamma$ such that $(A,\alpha) = (G,\gamma)[\iota]$.
\end{cor}
\proof
  Simply take $G = \SGlue\Phi\,A\,B\,f$; by
  Theorem~\ref{thm:fib-sglue} we get a composition structure for $G$, which
  we then adapt using Theorem \ref{thm:adapt} to get a new composition
  structure $\gamma$ satisfying the required equality.
\qed

\section{Univalence}
\label{univalence}

Voevodsky's \emph{univalence axiom}~\cite[Section~2.10]{HoTT} for a
universe $\mathcal{V}$ in a CwF (with at least $\Sigma$-, $\Pi$- and
$\Id$-types) states that for every $A, B : \mathcal{V}$ the canonical
function from $\Id \mathcal{V}\, A\, B$ to
$(f : A \fun B) \times \Equiv f$ is an equivalence.  Cohen~\emph{et
  al} construct a universe in the (CwF associated to the) presheaf
topos of cubical sets whose family of types is generic for CCHM
fibrations with small fibres (for a suitable notion of smallness in
the meta-theory) and prove that it satisfies the univalence
axiom. They do so by adapting the Hofmann-Streicher universe
construction for presheaf categories~\cite{HofmannM:lifgu}. It is not
possible to express their universe construction just using the
internal type theory of a general topos, for reasons that we discuss
below in Remark~\ref{rem:universes}. In recent work the authors, along
with Licata and Spitters \cite{LicataD:intumhtt}, have shown how to
axiomatize the CCHM universe construction in a modal extension of the
internal type theory. Here we just prove a version of univalence
without reference to a universe of fibrations.

To understand what this might mean, consider the following: were there
to be a universe $\UnivFib$ whose elements are codes for CCHM
fibrations, then given fibrations $(A,\alpha),(B,\beta):\Fib\Gamma$
named by functions $a,b : \Gamma \fun \UnivFib$ into the universe, a
path-equality between $a$ and $b$ gives (by Currying) a function
$p : \Gamma \times\I\fun \UnivFib$ such that $p(x,\src)=a\,x$ and
$p(x,\tgt)=b\,x$ for all $x:\Gamma$. Then $p$ names a fibration
$(P,\rho) : \Fib(\Gamma\times\I)$ such that
$(P,\rho)[\langle\id,\src\rangle] = (A,\alpha)$ and
$(P,\rho)[\langle\id,\tgt\rangle] = (B,\beta)$. The latter gives a
notion of path-equality between type-families whether or not there is
such a $\UnivFib$, which we study in this section in relation to
equivalences between fibrations.

To give the definitions more formally, we expand our running
assumption (see section~\ref{sec:intttt}) that the ambient topos $\E$
comes with a universe (internal full subtopos) $\Univ$ to the case
where there is a second universe $\Univ_1$ with $\Univ:\Univ_1$. We
sometimes refer to objects of type $\Univ$ as \emph{small} types and
objects of type $\Univ_1$ as \emph{large} types.

\begin{defi}[\textbf{Path equality between fibrations}]
  \label{defi:patebf}
  Define the type of paths between CCHM fibrations
  $\IdU{\_}{\_} : \{\Gamma:\Univ\} \fun \Fib\Gamma \fun \Fib\Gamma
  \fun \Univ_1$ by
  \[
  \IdU{A}{B} \defeq \{ P : \Fib(\Gamma\times\I) \mid
  P[\langle\id,\src\rangle] = A \conj
  P[\langle\id,\tgt\rangle] = B\}
  \]
\end{defi}

We show that given such a path $(P,\rho) : \IdU{(A,\alpha)}{(B,\beta)}$ it is always possible to
construct an equivalence $f : (x:\Gamma) \fun A\,x \fun B\,x$.
Conversely, given an equivalence $f : (x:\Gamma) \fun A\,x \fun B\,x$
between fibrations $(A,\alpha)$ and $(B,\beta)$, it is always possible
to construct such a $(P,\rho)$.

\begin{thm}[\textbf{Converting paths to equivalences}]
\label{thm:pathToEquiv}
  There is a function
  \begin{equation}
    \label{eq:42}\\
    {\pathToEquiv} : \{\Gamma:\Univ\}\{A\; B :\Fib\Gamma\}(P : \IdU{A}{B}) \\
    \fun (f : (x:\Gamma) \fun A\,x \fun B\,x) \times \Equiv f
  \end{equation}
\end{thm}
\proof 
Given
$\Gamma:\Univ, (A,\alpha), (B,\beta):\Fib\Gamma$ and $(P,\rho):\IdU{A}{B}$
we define maps $f : (x:\Gamma) \fun A\,x \fun B\,x$ and
$g : (x:\Gamma) \fun B\,x \fun A\,x$. First, given $x : \Gamma$ write
$\langle x,\id\rangle : (x,\src) \path (x,\tgt)$ for the path given by
$\langle x,\id\rangle\,i \defeq (x,i)$. Now define $f$ and $g$ as
follows:
\[
  f\,x\, a \defeq \rho\, \src\, \langle x,\id\rangle\, \False\,
  \elim_\emptyset\,a 
  \qquad\qquad 
  g\,x\, b \defeq \rho\, \tgt\, \langle x,\id\rangle\, \False\,
  \elim_\emptyset\, b 
\]
This definition is well-typed since $P(x,\src) = A\,x$ and
$P(x,\tgt) = B\,x$.  Since both functions are defined using
composition structure we can use filling (Lemma~\ref{lem:fill}) to
find dependently typed paths:
\begin{gather*}
  p : (x:\Gamma)(a : A\,x) \fun \Pi_\I P \text { defined by } p\, x\,
  a \defeq \fil 
  \src\, \rho\, \langle x,\id\rangle\, \False\, \elim_\emptyset\, a\\ 
  q : (x:\Gamma)(b : B\, x) \fun \Pi_\I P \text { defined by } q\, x\,
  b \defeq \fil 
  \tgt\, \rho\, \langle x,\id\rangle\, \False\, \elim_\emptyset\, b 
\end{gather*}
Note that for all $x:\Gamma$ and $a : A\,x$ we have
$p\, x\, a\, \src = a$ and $p\, x\, a\, \tgt = f\,x\, a$.  Similarly,
for all $b : B\,x$ we have $q\,x\, b\, \src = g\,x\, b$ and
$q\,x\, b\, \tgt = b$.  Now we define:
\begin{align*}
  r &: (x:\Gamma)(a : A\,x) \fun a \path g\,x\, (f\,x\, a)\\
  r &\, x\, a\, i \defeq \rho\, \tgt\, \langle x,\id\rangle\, (i=\src
      \disj i=\tgt)\, 
      ((\lambda\,\_ \fun p\,x\, a) \cup (\lambda\, \_ \fun q\,x\,
      (f\,x\, a)))\, 
      (f\,x\, a)\\ 
  s & : (x:\Gamma)(b : B\,x) \fun b \path f\,x\, (g\,x\, b)\\
  s &\, x\, b\, i \defeq \rho\, \src\, \langle x,\id\rangle\, (i=\src
      \disj i=\tgt)\, 
      ((\lambda\,\_ \fun q\,x\, b) \cup (\lambda\, \_ \fun p\,x\,
      (g\,x\, b)))\, 
      (g\,x\, b) 
\end{align*}
Hence $f$ and $g$ are quasi-inverses; from which we can construct an
equivalence structure~\cite[Chapter 4]{HoTT}. 
\qed

We now wish to show that, conversely, one can convert equivalences to
paths between fibrations. To do so we use the glueing construction
given in Section~\ref{sec:glueing}.

\begin{thm}[\textbf{Converting equivalences to paths}]
\label{thm:equivToPath}
  There is a function
  \begin{equation}
    \label{eq:14}
    {\equivToPath} : 
    \begin{array}[t]{@{}l}
      \{\Gamma:\Univ\}\{A\; B : \Fib\Gamma\}\\ 
      (f : (x:\Gamma)\fun \fst\,A\,x\fun \fst\,B\,x) \fun (\Equiv f) \fun
      \IdU{A}{B}
    \end{array}
  \end{equation}
\end{thm}
\proof 
Define the following:
\begin{align*}
  \Phi &: \Gamma\times\I \fun \Cof\\
  \Phi&\, (x,i) \defeq (i = 0) \disj (i = 1)\\
  C &: (\Gamma\times\I) \resby \Phi \fun \Univ\\
  C&\,((x,i),u) \defeq ((\lambda \_ : [i=0] \fun A\,x) \join (\lambda
     \_ : [i=1] 
     \fun B\,x))\, u\\ 
  f' &: ((x,i) : \Gamma\times\I)(u : [\Phi(x,i)]) \fun C((x,i),u) \fun B\, x\\
  f'&\,(x,i) \defeq (\lambda \_ : [i=0] \fun f\,x) \join (\lambda \_ :
      [i=1] 
      \fun id) 
\end{align*}
Now let $P \defeq \SGlue\, \Phi\, C\, (\lambda(x,\_) \fun B\,x)\, f'$
and observe that $P(x,\src) = A\,x$ and $P(x,\tgt) = B\,x$ by the
strictness 
property of $\SGlue$. 

Now we show that $P$ has a fibration structure.  First, we observe
that $C$ has a fibration structure, using axiom $\pos 1$. In order to
define $\gamma : \isFib\, C$ we take:
\[
  e : \Bool, \quad
  p : \I \fun (\Gamma\times\I)\resby\Phi, \quad
  \varphi : \Cof, \quad
  f : [ \varphi ] \fun \Pi (C \circ p), \quad
  c : C\, (p\, e)
\]
and aim to define $\gamma\,e\,p\,\varphi\,f\,c:C(p\,\inv{e})$. First, define
the predicate
$\kw{pZero} : \I \fun \Omega$ by:
\[
\kw{pZero}\, i \defeq (\snd (\fst (p\, i)) =  \src)
\]
Observe that since $p : \I \to (\Gamma\times\I)\resby\Phi$ we know,
from $\snd \comp\, p$, that: 
\begin{equation}
  \label{eq:allZeroOne} (\forall i : \I)\, \snd(\fst(p\, i)) = \src
  \disj \snd(\fst(p\, i)) = \tgt 	
\end{equation}
and so, using $\pos{2}$, we have
$(\forall i : \I)\, (\kw{pZero}\, i \disj \neg(\kw{pZero}\, i))$ and
using $\pos 1$ we get
\[
  ((\forall i : \I)\, \kw{pZero}\, i) \disj ((\forall i :
  \I)\,\neg(\kw{pZero}\, i)) 
\]
In case $(\forall i : \I)\, \kw{pZero}\, i$, we deduce that
$C \comp p = A \comp {\fst} \comp {\fst} \comp p$ and define:
\[ 
\gamma\, e\, p\, \varphi\, f\, c \defeq \alpha\, e\, ({\fst} \comp
{\fst} \comp p)\, \varphi\, f\, c  
\] 
Otherwise, in case $(\forall i : \I)\,\neg(\kw{pZero}\, i)$, we use
$\pos{2}$ to deduce
$(\forall i : \I)\, \snd(\fst(p\, i)) = \tgt$, hence
$C \comp p = B \comp {\fst} \comp {\fst} \comp p$, and so define:
\[ 
\gamma\, e\, p\, \varphi\, f\, c \defeq \beta\, e\, ({\fst} \comp
{\fst} \comp p)\, \varphi\, f\, c  
\] 
Therefore $C$ has a fibration structure. Next we show that $f'$ is an
equivalence for every $x:\Gamma$, $i : \I$ and $u : [\Phi\, i]$.
First note that the identity function $\id : B\,x \fun B\,x$ is always an
equivalence; let $idEq : \Equiv\, \id$ be a proof of this
fact. Define $eq' : (x:\Gamma)(i : \I)(u : [ \Phi\, i ]) \fun \Equiv\,
(f'\,x\, i\, u)$ by:
\[
  eq'x\,\,i \defeq (\lambda u:[i=\src] \fun eq) \cup (\lambda
  u:[i=\tgt] \fun idEq) 
\]
Hence, by Corollary 
\ref{cor:glue}, we get a fibration structure, $\rho : \isFib P$, such that
\[(P,\rho)[\iota : (\Gamma\times\I)\resby\Phi \mono \Gamma\times\I ] =
(C,\gamma)\]
and hence 
$(P,\rho)[\langle\id,\src\rangle] = (P,\rho)[\iota \comp
\langle\id,\src,*\rangle] = (C,\gamma)[\langle\id,\src,*\rangle] =
(A,\alpha)$ 
and similarly
$(P,\rho)[\langle\id,\tgt\rangle] = (B,\beta)$.
Therefore we define
$\equivToPath\, f\, e \defeq (P, \rho)$. \qed

The univalence axiom~\cite[Axiom 2.10.3]{HoTT} becomes here the
property that the map ${\pathToEquiv}$ is itself an equivalence. A
result by Licata [unpublished] tells us that this is actually
the same as having a map from equivalences to paths, which we have
in ${\equivToPath}$, such that coercion along $\equivToPath\,f\,e$ is
path equal to $f$. This property does indeed hold:

\begin{thm}[\textbf{Univalence for $\IdU{}{}$}]
  \label{thm:uni}
  Define 
  \begin{align*}
  &\coerce : \{\Gamma:\Univ\}\{A\; B :\Fib\Gamma\}(P :
  \IdU{A}{B})(x : \Gamma) \fun \fst\,A\, x \fun \fst\,B\,x \\
  &\coerce (P,\rho) \defeq \fst(\pathToEquiv(P,\rho)) 
  \end{align*}
  Given $\Gamma:\Univ$, $(A,\alpha),(B,\beta):\Fib\Gamma$,
  $f:(x:\Gamma)\fun A\,x\fun B\,x$ and $eq:\Equiv\,f$, there exists a
  path $f \path \coerce(\equivToPath\,f\,e)$.
\end{thm}
\proof 
Let $(P,\rho) \defeq \equivToPath\,f\,e$. Unfolding the definition of
$\coerce$ we have
\[ 
\coerce\,(P,\rho)\,x\, a = \rho\, \src\, \langle x,\id\rangle\,
\False\, \elim_\emptyset\,a
\]
Recalling that $\rho$ is the composition structure for $\SGlue$ we can
calculate
$\rho\, \src\, \langle x,\id\rangle\, \False\, \elim_\emptyset\,a$. We
first move across the isomorphism with $\Glue$ so that $a$ becomes
$(\lambda\_\fun a,f\,x\,a)$. We now trace the algorithm for composition in
$\Glue$:
we begin by composing in $B\,x$ to get
$b'_1 \defeq \beta\,\src\,(\lambda\_\fun x)\,\False\,
\elim_\emptyset\,(f\,x\,a)$.
Next, we use the equivalence structure to derive
$a_1 : (u:[\tgt=\src\disj\tgt=\tgt])\fun B\,x$ and
$p_b : (u:[\tgt=\src\disj\tgt=\tgt])\fun \id\, (a_1\,u) \path b'_1$, where
in particular, $a_1$ is given by $a_1 u\defeq \beta\,\tgt\,(\lambda\_\fun
x)\,\False\, \elim_\emptyset\,b'_1$. We then
 perform the final step of the composition, which is to compose from $b'_1$ in
 $B\,x$ to get $b_1$. This leaves us with the result of composing in $\Glue$ as $(a_1, b_1)$. When transferring back across the  isomorphism we simply take the first component of the pair, namely $a_1$, but
now regarded as a total element to get the final result of composing in
$\SGlue$ as $\beta\,\tgt\,(\lambda\_\fun
x)\,\False\, \elim_\emptyset\,b'_1$. So, in summary we have:
\begin{align*}
\coerce\,(P,\rho)\,x\, a &= \rho\, \src\, \langle x,\id\rangle\,
\False\, \elim_\emptyset\,a \\
&= \beta\,\tgt\,(\lambda\_\fun
x)\,\False\, \elim_\emptyset\,b'_1\\
&= \beta\,\tgt\,(\lambda\_\fun
x)\,\False\, \elim_\emptyset\, (\beta\,\src\,(\lambda\_\fun
x)\,\False\, \elim_\emptyset\,(f\,x\,a))
\end{align*}
Since this is simply two trivial compositions applied to $f\, x\, a$ we can
use use filling to derive a path $f\,x\,a \path \coerce\,(P,\rho)\,x\, a$.
Now, two applications of function extensionality
(Remark~\ref{rem:funext}) yields a path $f \path \coerce\,(P,\rho)$
 as required.  \qed

\begin{rem}
  \label{rem:universes}
  Suppose we have a ``Tarski-style'' universe in the CwF of CCHM
  fibrations (Definition~\ref{def:cchm-fib-cwf}), that is, a (large)
  fibrant object $\UnivFib:\Univ_1$ and a fibration
  $(\El, \eta) : \Fib{\UnivFib}$ (with small fibres,
  $\El:\UnivFib\fun\Univ$). We can regard the elements $a:\UnivFib$ as
  \emph{codes} for small types $\El\,a : \Univ$ and ask that the
  universe have functions on codes for type-forming constructs of
  interest, such as $\Sigma$, $\Pi$ and $\Id$;
  \emph{cf.}~\cite[Section~2.1.6]{HofmannM:synsdt}).  Re-indexing
  $(\El, \eta)$ along a function $a:\Gamma\fun\UnivFib$ gives a
  fibration $(\El \comp a, \eta[a]): \Fib\Gamma$; and re-indexing
  turns paths in $\UnivFib$ into path equalities between fibrations in
  the sense of Definition~\ref{defi:patebf}. In order to use
  Theorem~\ref{thm:uni} to deduce that such a universe is univalent,
  one would at least need to construct functions
  \[
  \IdU{(\El \comp a, \eta[a])}{(\El \comp b, \eta[b])} \fun (a\path b) 
  \qquad(\text{for\ } a,b:\Gamma\fun\UnivFib)
  \]
  converting path equalities between fibrations arising from codes
  back into paths between the codes themselves. For the universe in
  the presheaf topos of cubical sets constructed by Cohen \emph{el
    al}~\cite[Definition~18]{CoquandT:cubttc}, the existence of such
  functions with good properties follows from the fact that the
  universe is generic, in the sense that for every object
  $\Gamma$ and every CCHM fibration $(A,\alpha)$ over $\Gamma$, there
  is a morphism $a=\code{A}{\alpha}:\Gamma\morphism \UnivFib$ with
  $(A,\alpha)=(\El \comp a, \eta[a])$. One difficulty for our approach
  of using the internal type-theoretic language of a topos is that
  this genericity is an external, or global, property:
  postulating an internal version of it, i.e.~the existence of a function
  \[\code{\_}{\_} : \{\Gamma : \Univ_1\}(A : \Gamma \fun \Univ_1)
      (\alpha : \isFib{A}) \fun \Gamma \fun \UnivFib\]
  such that
  \[(\El \comp\code{A}{\alpha}, \eta[\code{A}{\alpha}]) = (A,\alpha)\]
  leads to contradiction. To see why, recall Remark \ref{rem:fibo}
  which points out that a family of fibrant objects is not necessarily
  fibrant. Were we to assume the existence of such a map
  $\code{\_}{\_}$, then we could derive a fibration structure for a
  family of fibrant objects like so: given an object $\Gamma:\Univ$,
  and family $A:\Gamma\fun\Univ$ write $A_x : 1\fun\Univ$, where
  $x : \Gamma$, for the object $A\;x$ as a family over $1$,
  i.e.~$A_x \defeq \lambda(\_:1) \fun A\, x$.  If $A$ has fibrant
  fibres then we have a map $\alpha_{\_} : (x:\Gamma) \fun \Fib{A_x}$,
  so that $\alpha_x : \Fib{A_x}$ is a fibration structure for the
  object $A\;x$ regarded as a family over $1$ (for $x : \Gamma$). Then
  we can define a function $a : \Gamma\fun\UnivFib$ as follows:
  \[ 
  a\, x \defeq \code{A_x}{\alpha_x}(*) 
  \]
  (where $*$ is the unique element of the terminal type $1$). It can
  be checked that $El \comp a = A$ and therefore $\eta[a] : \Fib{A}$.
  Hence we have a fibration structure for $A$ whenever $A$ has fibrant
  fibres; but this gives a contradiction as in Remark~\ref{rem:fibo}.
  This issue can be resolved by moving to a different internal type
  theory where one has access to a modal operator which is able to
  express properties of global, rather than local, elements
  \cite{LicataD:intumhtt}.
\end{rem}

\section{Satisfying  the Axioms}
\label{sec:sata}

Working informally in a constructive set theory, the authors of
\cite{CoquandT:cubttc} give a model of their type theory using the
topos $\hat{\dm}=\Set^{\dm^\op}$ of contravariant set-valued functors
on a particular small category $\dm$ that they call the \emph{category
  of cubes}. In this section we present sufficient conditions on an
arbitrary small category $\C$ for the topos $\hat{\C}=\Set^{\C^\op}$
of set-valued presheaves (within Intuitionistic ZF set
theory~\cite[Section~3.2]{AczelP:notcst}, say) to have an interval
object and subobject of cofibrant propositions satisfying the axioms
in Figure~\ref{fig:axi}. We show that the category of cubes is an
instance of such a $\C$ (in more than one way); and we also show that
in the presence of the Law of Excluded Middle, so is the simplex
category $\Simplex$.

We begin by briefly recalling the definition of $\dm$ from
\cite[Section~8]{CoquandT:cubttc}.  Recall that a \emph{De~Morgan
  algebra} is a distributive lattice equipped with a function
$d\mapsto \Not{d}$ which is involutive $\Not{(\Not{d})} = d$ and
satisfies De~Morgan's Law
$\Not{(d_1 \disj d_2)} = (\Not{d_1})\conj(\Not{d_2})$; see
\cite[Chapter~XI]{BalbesR:disl}. A homomorphism of De~Morgan algebras
is a function preserving finite meets and joins and the involution
function. Let $\DM$ denote the category of De~Morgan algebras and
homomorphisms and $\fdm(I)$ the free De~Morgan algebra on a set $I$.

\begin{defi}[\textbf{The category of cubes, $\dm$}]
  \label{defi:catc}
  Fix a countably infinite set $\D$ whose elements we call
  \emph{directions} and write as $\d,\d[y], \d[z],\ldots$  The
  objects of $\dm$ are the finite subsets of $\D$, which we write as
  $I,J,K,\ldots$ The morphisms $\dm(I,J)$ are all functions
  $J\morphism \fdm(I)$. Such functions are in bijection with the
  De~Morgan algebra homomorphisms $\fdm(J)\morphism \fdm(I)$ and the
  composition in $\dm$ of $f\in\dm(I,J)$ with $g\in\dm(J,K)$ is the
  composition in $\Set$ of $g:K\morphism \fdm(J)$ with the
  homomorphism $\fdm(J)\morphism\fdm(I)$ corresponding to $f$.
\end{defi}

Thus $\dm$ is equivalent to the opposite of the full subcategory of
$\DM$ whose objects are the free, finitely generated De~Morgan
algebras. Hence it is the algebraic theory of De~Morgan algebra as
a Lawvere theory~\cite{LawvereFW:funsat,AdamekJ:algtci}: it is a
category with finite products equipped with an internal De~Morgan
algebra (whose underlying object is $\{\d\}$ for some chosen
$\d\in\D$) and is universal among such categories.

\subsection{The interval object
  (\texorpdfstring{$\pos{1}$}{ax1}--\texorpdfstring{$\pos{4}$}{ax4})}
\label{subsec:into}

Returning to the general case of a small category $\C$ with its
associated presheaf topos $\hat{\C}$, if we take the interval object
$\I \in \hat{\C}$ to be a representable functor
$\y_{\iobj} \defeq \C({\_}\mathbin{,}\iobj)$ for some object
$\iobj\in\C$, then the following theorem gives a useful criterion for
such an interval object to satisfy axiom $\pos{1}$.

\begin{thm}
  \label{thm:sat-pos1}
  In a presheaf topos $\hat{\C}$, a representable functor
  $\I=\y_{\iobj}$ satisfies axiom $\pos{1}$ if $\C$ is a
  \emph{cosifted} category, that is, if finite products in $\Set$
  commute with colimits over $\C^{\op}$~\cite{GabrielP:lokpk}.
\end{thm}
\proof
  $\C$ is cosifted if the colimit functor
  $\colim_{\C^{\op}}:\hat{\C}\fun\Set$ preserves finite
  products. Recall that $\colim_{\C^{\op}}:\hat{\C}\fun\Set$ is left
  adjoint to the constant presheaf functor $\Delta:\Set\fun \hat{\C}$
  and (hence) that for any $c\in\C$ it is the case that
  $\colim_{\C^{\op}}\y_c \iso 1$. So when $\C$ is cosifted we have for
  any $c\in\C$
  \begin{multline*}
    \hat{\C}(\y_c \times \y_{\iobj},\Delta\{0,1\}) \iso
    \Set(\colim_{\C^{\op}}(\y_c \times \y_{\iobj}), \{0,1\}) \iso {}\\
    \Set(\colim_{\C^{\op}}\y_c \times \colim_{\C^{\op}}\y_{\iobj},
    \{0,1\}) \iso \Set(1\times 1,\{0,1\}) \iso \{0,1\}
  \end{multline*}
  Since decidable subobjects in $\hat{\C}$ are classified by
  $1+1=\Delta\{0,1\}$, this means that the only two decidable
  subobjects of $\y_c \times \y_{\iobj}$ are the smallest and the
  greatest subobjects. Since this is so for all $c\in\C$, it follows
  that $\I=\y_{\iobj}$ satisfies $\pos{1}$. 
\qed

A more elementary characterisation of cosiftedness is that $\C$ is
inhabited and for every pair of objects $c,c'\in\C$ the category of
spans $c\leftarrow\cdot\rightarrow c'$ is a connected
category~\cite[Theorem~2.15]{AdamekJ:algtci}. Any category with finite
products trivially has this property. This is the case for the
category $\dm$ of cubes defined above and thus the interval in the
model of~\cite{CoquandT:cubttc} (where $\C=\dm$ and $\iobj$ is the
generic De~Morgan algebra) satisfies $\pos{1}$. A relevant example of
a category that does not have finite products, but which is
nevertheless cosifted is $\Simplex$, the category of inhabited finite
linearly ordered sets $[0<1<\cdots<n]$, for which $\hat{\C}$
is the category of \emph{simplicial sets}, widely used in homotopy
theory~\cite{GoerssP:simhtp}. Thus the natural candidate for an
interval in $\hat{\Simplex}$, namely $\y_{\iobj}$ when $\iobj$ is the 1-simplex
$[0<1]$, satisfies $\pos{1}$.

In addition to $\pos{1}$, the other axioms in Figure~\ref{fig:axi}
concerning the interval say that $\I$ is a non-trivial ($\pos{2}$)
model of the algebraic theory given by $\pos{3}$ and $\pos{4}$, which
we call \emph{connection algebra}. (See also Definition~1.7 of
\cite{GambinoN:frocrp}, which considers a similar notion in a more
abstract setting.) For cubical sets, the Yoneda embedding
$\y:\dm\morphism \hat{\dm}$ sends the generic De~Morgan algebra in
$\dm$ to a De~Morgan algebra in $\hat{\dm}$.  This is a non-trivial
connection algebra: the constants are the least and greatest
elements and the binary operations are meet and join. An obvious
variation on the theme of~\cite{CoquandT:cubttc} would be to replace
$\dm$ by the Lawvere theory for connection algebras. Note also
that the 1-simplex in $\sSet$ is a non-trivial connection
algebra, the constants being its two end points and the binary
operations being induced by the order-preserving binary operations of
minimum and maximum on $[0<1]$.

\subsection{Cofibrant propositions
  (\texorpdfstring{$\pos{5}$}{ax5}--\texorpdfstring{$\pos{8}$}{ax8})
  and the strictness axiom
  (\texorpdfstring{$\pos{9}$}{ax9})} \label{sec:sat-strictness} 
 
In a topos with an interval object, there are many candidates for a
subobject $\Cof\mono\Prop$ satisfying axioms $\pos{5}$--$\pos{8}$ in
Figure~\ref{fig:axi}. At one extreme, one could just take $\Cof$ to be
the whole of $\Omega$. At the opposite extreme, one could take the
subobject (internally) inductively defined by the requirements that it
contains the propositions $i=\src$ and $i=\tgt$ (for all $i:\I$) and
is closed under binary disjunction, dependent and $\I$-indexed
conjunction, thereby obtaining the smallest $\Cof$ satisfying
$\pos{5}$--$\pos{8}$. However, cofibrant propositions also have to
satisfy the strictness axiom $\pos{9}$ and we consider that next.

Given a presheaf topos $\hat{\C}$, we work in the
CwF associated with $\hat{\C}$ as in
\cite[Section~4]{HofmannM:synsdt}. In particular, families over a
presheaf $\Gamma\in\hat{\C}$ are given by functors
$(\int \Gamma)^{\op}\fun \Set$, where $\int\Gamma$ is the usual
category of elements of $\Gamma$, with
$\obj(\int \Gamma) = (c\in \obj\C)\times \Gamma\,c$ and
$(\int\Gamma)((c,x),(d,y)) = \{f\in \C(c,d) \mid \Gamma\,f\,y = x\}$.
If $\S$ is a Grothendieck universe in the ambient set theory, then its
Hofmann-Streicher lifting~\cite{HofmannM:lifgu} to a universe $\Univ$
in that CwF satisfies that the morphisms $\Gamma\fun\Univ$ in
$\hat{\C}$ name the families $(\int \Gamma)^{\op}\fun \S$ taking
values in $\S\subseteq\Set$.

\begin{defi}[$\Dec$]
  \label{defi:omed}
  The subobject classifier $\Prop$ in a presheaf topos $\hat{\C}$ maps
  each $c\in\C$ to the set $\Omega(c)$ of \emph{sieves} on $c$, that
  is, pre-composition closed subsets $S\subseteq\obj(\C/c)$. Let
  $\Dec\mono\Prop$ be the subpresheaf whose value at each $c\in\obj\C$
  is the subset of $\Omega(c)$ consisting of those sieves $S$ that are
  decidable subsets of $\obj(\C/c)$.
\end{defi}

Of course if the ambient set theory satisfies the Law of Excluded
Middle, then $\Dec=\Prop$. In general $\Dec$ classifies monomorphisms
$\alpha:F\mono G$ in $\hat{\C}$ such that for all $c\in\obj\C$ the
(injective) function $\alpha_c:F\,c\fun G\,c$ has decidable image.

\begin{thm}
  \label{thm:strictness}
  Interpreting the universe $\Univ$ as the Hofmann-Streicher
  lifting~\cite{HofmannM:lifgu} of a Grothendieck universe in $\Set$,
  a subobject $\Cof\mono\Prop$ in a presheaf topos $\hat{\C}$
  satisfies the strictness axiom $\pos{9}$ if it is contained in
  $\Dec\mono\Prop$.
\end{thm}
\proof
  For each $c\in\obj\C$, suppose we are given $S\in\Dec(c)$. Thus $S$
  is a sieve on $c$ and for each $c'\in\obj\C$ and $\C$-morphism
  $f:c'\fun c$, it is decidable whether or not $f\in S$. We can also
  regard $S$ as a subpresheaf $S\hookrightarrow\y_c$.

  Suppose that we have families $A:(\int S)^{\op}\fun \S$,
  $B:(\int \y_{c})^{\op}\fun \S$ and a natural isomorphism $s$ between
  $A$ and the restriction of $B$ along $S\hookrightarrow \y_c$.  For
  each $\C$-morphism $\cdot\xrightarrow{f}c$, using the decidability
  of $S$, we can define bijections $s'(f):B'(f)\iso B(f)$ given by
  \[
  B'(f) \defeq
  \begin{cases}
    A(f) &\text{if $f\in S$}\\
    B(f) &\text{if $\neg(f\in S)$}
  \end{cases}
  \qquad\text{and}\qquad
  s'(f) \defeq
  \begin{cases}
    s(f) &\text{if $f\in S$}\\
    f &\text{if $\neg(f\in S)$}
  \end{cases}
  \]
  (compare this with Definition~15 in~\cite{CoquandT:cubttc}).
  We make $B'$ into a functor ${(\int \y_{c})}^{\op}\fun \S$ by
  transferring the functorial action of $B$ across these
  bijections. Having done that, $s'$ becomes a natural isomorphism
  $B'\iso B$; and by definition its restriction along $S\hookrightarrow
  \y_c$ is $s$. 
\qed

\begin{cor}
  \label{cor:cart}
  Let $\C$ be a small category with finite products containing an
  object $\iobj$ with the structure of a non-trivial connection
  algebra $\src,\tgt:1\fun\iobj$,
  ${\sqcap},{\sqcup}:\iobj\times \iobj \fun \iobj$
  (cf.~Figure~\ref{fig:axi}).  Suppose that for each object $c\in\C$,
  the set $\C(c,\iobj)$ has decidable equality.  Then the topos of
  presheaves $\hat{\C}$ satisfies all the axioms in
  Figure~\ref{fig:axi} if we take the interval $\I$ to be $\y_\iobj$
  and $\Cof$ to be $\Dec$.
\end{cor}
\proof We already noted in section~\ref{subsec:into} that axioms
$\pos{1}$--$\pos{4}$ are satisfied by this choice of $\I$.  The
decidability of each set $\C(c,\iobj)$ implies that the subobjects
$\{\src\}\mono\I$ and $\{\tgt\}\mono\I$ in $\hat{\C}$ factor through
$\Dec\mono\Omega$ and hence that axiom $\pos{5}$ is satisfied when
$\Cof=\Dec$.  Note that this choice of $\Cof$ automatically satisfies
$\pos{6}$ and $\pos{7}$; and it satisfies axiom $\pos{9}$ by
Theorem~\ref{thm:strictness}. So it just remains to check that axiom
$\pos{8}$ is satisfied. We saw in Lemma~\ref{lem:ax7-8}(ii) that this
axiom is equivalent to requiring cofibrations to be closed under
exponentiation by $\I$.  In this case cofibrations are the
monomorphisms $\alpha:F\mono G$ classified by $\Dec$ and we noted
after Definition~\ref{defi:omed} that they are characterized by the
fact that each function $\alpha_c\in\Set(F\,c, G\,c)$ has decidable
image. Closure of these monomorphisms under exponentiating by $\I$
follows from the fact that $\C$ has finite products and that
$\I=\y_{\iobj}$ is representable; for then $\I\fun(\_)$ is isomorphic
to the functor $\hat{\C}\morphism\hat{\C}$ induced by precomposition
with $(\_)\times\iobj:\C\morphism\C$, which clearly preserves the
componentwise decidable image property. \qed

The above argument for axiom~$\pos{8}$ does not apply to $\Simplex$,
since it does not have finite products; we do not know whether
that axiom is satisfied by constructive simplicial sets. However, in
the presence of the Law of Excluded Middle (LEM), $\Dec=\Omega$ and we
have:

\begin{cor}[Classical simplicial sets]
  Assuming LEM holds in the set-theoretic meta-theory, then the
  presheaf topos of simplicial sets $\hat{\Simplex}$ satisfies the
  axioms in Figure~\ref{fig:axi} if we take $\I$ to be the
  representable presheaf on the 1-simplex and $\Cof$ to be the whole
  of $\Omega$.
\end{cor}
\proof
  We already noted in section~\ref{subsec:into} that axioms
  $\pos{1}$--$\pos{4}$ are satisfied by this choice of $\I$. If
  $\Cof=\Omega$ (that is, $\cof=\lambda\_\fun\True$), then axioms
  $\pos{5}$--$\pos{8}$ hold trivially. Furthermore, if LEM holds, then
  $\Dec=\Omega$ and so axiom~$\pos{9}$ holds by
  Theorem~\ref{thm:strictness}.
\qed

\begin{rem}
  As a partial converse of Theorem~\ref{thm:strictness}, we have that
  if $\pos{9}$ is satisfied by the Hofmann-Streicher universe in the
  CwF associated with $\hat{\C}$, then each cofibrant mono
  $\alpha:F\mono G$ has component functions
  $\alpha_c\in\Set(F\,c,G\,c)$ ($c\in\obj\C$) whose images are
  $\neg\neg$-closed subsets of $G\,c$. To see this we can apply an
  argument due to Andrew Swan [private communication] that relies upon
  the fact that in the ambient set theory one has
  \begin{equation}
    \label{eq:19}
    (X=\emptyset) \;=\; \forall x\in X.\;\False \;=\;
    \neg\neg(\forall x\in X.\;\False) \;=\; \neg\neg(X=\emptyset)
  \end{equation}
  For suppose given $c\in\obj\C$ and $S\in\Cof(c)$. We have to use
  axiom $\pos{9}$ to show that $S$ is a $\neg\neg$-closed subset
  of $\obj(\C/c)$.  Let $A:{(\int S)}^{\op}\fun \S$ be the constant
  functor mapping each $(c',f)$ to $\{\emptyset\}$; and let
  $B:{(\int \y_{c})}^{\op}\fun \S$ map each $(c',f)$ to
  $\{\{\emptyset\},\{\emptyset\mid f\in S\}\}$ (which does extend to a
  functor, because $S$ is a sieve). The restriction of $B$ along
  $S\hookrightarrow \y_c$ is isomorphic to $A$ and so by $\pos{9}$
  there is some $B': {(\int \y_{c})}^{\op}\fun \S$ whose restriction
  along $S\hookrightarrow \y_c$ is equal to $A$ and some isomorphism
  $s':B'\iso B$. For any $(c',f)\in\obj(\int \y_{c})$, suppose
  $X\in B'(c',f)$; then $f\in S \imp X=\emptyset$, hence
  $\neg\neg(f\in S) \imp \neg\neg(X=\emptyset)$ and therefore
  by~\eqref{eq:19}, $\neg\neg(f\in S) \imp (X=\emptyset)$.
  Therefore
  $\neg\neg(f\in S) \imp B'(c',f) =\{\emptyset\} \imp B(c',f) \iso
  \{\emptyset\} \imp {f\in S}$.
  So $S$ is indeed a $\neg\neg$-closed subset of $\obj(\C/c)$.

  Note that this result implies that it is not possible to take $\Cof$
  to be the whole of $\Omega$ and satisfy $\pos{9}$ unless the ambient
  set theory satisfies LEM.
\end{rem}

Since in a constructive setting equality in free De~Morgan algebras is
decidable, it follows that Corollary~\ref{cor:cart} gives a model of
our axioms when $\C=\dm$, the category of cubes
(Definition~\ref{defi:catc}). However, this uses a different choice of
cofibrancy from the one in~\cite[Section~4.1]{{CoquandT:cubttc}}. In
the remainder of this section we check that the CCHM notion of
fibrancy satisfies our axioms.

\begin{defi}[\textbf{Cofibrant propositions in~\cite{CoquandT:cubttc}}]  
  For each object of the category of cubes $I\in\obj\dm$, Cohen
  \emph{et al} define the \emph{face lattice} $\Face(I)$ to be the
  distributive lattice generated by symbols $(\d=\src)$ and
  $(\d=\tgt)$ for each $\d\in I$, subject to the equations
  $(\d=\src) \conj (\d=\tgt) = \bot$.
  
  Since the free De~Morgan algebra $\fdm(I)$ is freely generated as a
  distributive lattice by symbols $\d$ and $\Not{\d}$ (as $\d$ ranges
  over the finite set $I$), we can regard $\Face(I)$ as a quotient
  lattice of $\fdm(I)$ via the function mapping $\d$ to $(\d=\tgt)$
  and $\Not{\d}$ to $(\d=\src)$; we write
  $q_I : \fdm(I) \morphism \Face(I)$ for the quotient
  function.\footnote{If $r\in\fdm(I)$, then \cite{CoquandT:cubttc}
    uses the notation $(r=\tgt)$ for $q_I(r)$.} It is not hard to see
  that for each $f\in\C(J,I)$, the corresponding De~Morgan algebra
  homomorphism $\fdm(I)\morphism\fdm(J)$ (which we also write as $f$)
  induces a lattice homomorphism between the face lattices:
  \[
  \xymatrix{
    {\fdm(I)} \ar[r]^f \ar[d]_{q_I} & 
    {\fdm(J)} \ar[d]^{q_J}\\
    {\Face(I)} \ar[r]_{\Face\,f} &
    {\Face(J)}}
  \]
  This makes $\Face$ into an object of $\hat{\dm}$ and there is a
  monomorphism $m:\Face\mono\Omega$ whose component at $I\in\obj\dm$
  sends each $\varphi\in\Face(I)$ to the sieve
  \begin{equation}
    \label{eq:33}
    m_I(\varphi) = \{\cdot\xrightarrow{f} I \mid
    \Face\,f\,\varphi=\top\}  
  \end{equation}
\end{defi}

We now take $\Cof\mono\Omega$ in $\hat{\dm}$ to be the subobject given
by this monomorphism $m:\Face\mono\Omega$. Axioms
$\pos{1}$--$\pos{4}$ hold without change from
Corollary~\ref{cor:cart}; and axioms $\pos{5}$--$\pos{7}$ follow from
the definition of the face lattices $\Face(I)$. So it just remains to
check axioms $\pos{8}$ and $\pos{9}$.

Recall that the interval object in $\hat{\dm}$ is the representable
presheaf $\I=\y_{\{\d\}}$ on a one-element subset $\{\d\}\in\obj\dm$.
For an arbitrary object $I\in\obj\dm$, with $n$ distinct elements
$\d_1,\ldots\d_n\in\D$ say, the representable $\y_I$ is isomorphic to
an $n$-fold product $\I^n$ in $\hat{\dm}$. Thus the sieve
\eqref{eq:33} corresponds to a subobject of the $n$-cube
$\I^n$. Indeed, each $\varphi$ in the face lattice $\Face(I)$ is a
finite join of irreducible elements; and each of those irreducibles is
a finite conjunction of atomic conditions of the form $\d=\src$ or
$\d[y]=\tgt$. The corresponding subobject of $\I^n$ is a finite union
of \emph{faces}, that is, subobjects of $\I^n$ obtained by setting
some co-ordinates to either $\src$ or $\tgt$. This disjunctive normal
form for elements of $\Face(I)$ entails that its equality is decidable
and hence this $\Cof$ is contained in $\Dec$ and axiom $\pos{9}$ is
satisfied (Theorem~\ref{thm:strictness}). Finally, axiom $\pos{8}$
follows from the fact that pullback of cofibrations along a projection
$\pi_1:\I^n\times\I\morphism\I^n$ has a (stable) right adjoint, or
equivalently that the lattice morphisms
$\Face\,\pi_1:\Face(I) \morphism \Face(I\cup\{\d\})$ (for any
$I\in\obj\dm$ and $\d\in\D-I$) have (stable) right adjoints. This is
the quantifier elimination result for face lattices; see
\cite[Lemma~2]{CoquandT:cubttc}.

\section{Related Work}
\label{sec:relfw}

The work reported here was inspired by the suggestion of
Coquand~\cite{CoquandT:cork} that some of the constructions developed
in~\cite{CoquandT:cubttc} paper might be better understood using the
internal logic of a topos. We have shown how to express Cohen,
Coquand, Huber and M\"ortberg's notion of fibration in the internal
type theory of a topos. The use of internal language permits an
appealingly simple description (Definition~\ref{def:cchm-fib})
compared, for example, with the more abstract category-theoretic
methods of weak factorization systems and model categories, which have
used for the same purpose by Gambino and
Sattler~\cite[Section~3]{GambinoN:frocrp}. Birkedal \emph{et
  al}~\cite{BirkedalL:guactt} develop guarded cubical type theory with
a semantics based on an axiomatic version of~\cite{CoquandT:cubttc}
within the internal logic of a presheaf topos.

Within the framework of a topos equipped with an interval-like object,
we found that quite a simple collection of axioms
(Figure~\ref{fig:axi}) suffices for this to model Martin-L\"of type
theory with intensional identity types satisfying a weak form of
univalence. In particular, only a simple connection algebra, rather
than a de~Morgan algebra structure, is needed on the
interval. Furthermore, the collection of propositions suitable for
uniform Kan filling is not tightly constrained and can be chosen in
various ways. In Section~\ref{sec:sata} we only considered how
presheaf categories can satisfy our axioms. It might be interesting to
consider models in general Grothendieck toposes (where presheaves are
restricted to be sheaves for a given notion of covering), particularly
\emph{gros toposes} such as Johnstone's topological
topos~\cite{JohnstonePT:topt}; this allows the interval object to be
(a representable sheaf corresponding to) the usual topological
interval and hence for the model of type theory to have a rather
direct connection with classical homotopy types of spaces.  However,
although the Hofmann-Streicher~\cite{HofmannM:lifgu} universe
construction (the basis for the construction in Section~8.2
of~\cite{CoquandT:cubttc} of a fibrant universe satisfying the full
univalence axiom) can be extended from presheaf to sheaf toposes via
the use of sheafification~\cite[Section~3]{StreicherT:unit}, it seems
that sheafification does not interact well with the CCHM notion of
fibration. In another direction, recent work of Frumin and
Van~Den~Berg~\cite{FruminD:homtmf} makes use of our elementary,
axiomatic approach using a non-Grothendieck topos, namely the
effective topos~\cite{HylandJME:efft}.

\section*{Acknowledgements}

We thank Thierry Coquand for many conversations about the results
described in~\cite{CoquandT:cubttc} and the possibility of an internal
characterisation of its constructions. We are grateful to him,
Mart\'{\i}n Escard\'o, Anders M\"ortberg, Christian Sattler, Bas
Spitters, Andrew Swan and the anonymous referees for their comments on
the work. This paper is a revised and expanded version of a paper of
the same name that appeared in the proceedings of the 25th EACSL
Annual Conference on Computer Science Logic (CSL 2016).


\newcommand{\etalchar}[1]{$^{#1}$}

\end{document}